\documentclass[format=acmsmall, review=False, screen=true]{acmart}

\usepackage{booktabs} 
\usepackage{lipsum} 
\usepackage{balance}
\usepackage{paralist}
\usepackage{filecontents}
\usepackage{etoolbox}
\usepackage{mathrsfs}
\usepackage{bm}
\usepackage{setspace}
\usepackage{graphicx}
\usepackage{caption}
\usepackage{subcaption}
\usepackage[ruled]{algorithm2e}
\usepackage{multirow}
\usepackage{amsmath}
\usepackage{algpseudocode}
\usepackage{hyperref}
\usepackage{afterpage}

\DeclareMathOperator*{\argmin}{arg\,min}

\newcommand{\revised}[1]{\textcolor{black}{{\bf}{#1}{\bf}}}

\newtheorem{theorem}{Theorem}[section]
\newtheorem{definition}{Definition}[section]

\acmJournal{TOIS}
\acmVolume{9}
\acmNumber{4}
\acmArticle{29}
\acmYear{2020}
\acmMonth{12}
\copyrightyear{2020}

\setcopyright{acmlicensed}

\acmDOI{0000001.0000001}

\begin{document}
\title{Unbiased Learning to Rank: Online or Offline?}

\author{Qingyao Ai}
\affiliation{%
	\institution{School of Computing, University of Utah}
	\city{Salt Lake City} 
	\state{UT} 
	\country{USA}
	\postcode{84112-9205}
}
\email{aiqy@cs.utah.edu}

\author{Tao Yang}
\affiliation{%
	\institution{School of Computing, University of Utah}
	\city{Salt Lake City} 
	\state{UT} 
	\country{USA}
	\postcode{84112-9205}
}
\email{taoyang@cs.utah.edu}

\author{Huazheng Wang}
\affiliation{%
	\institution{Department of Computer Science, University of Virginia}
	\city{Charlottesville} 
	\state{VA} 
	\country{USA}
	\postcode{22904}
}
\email{hw7ww@virginia.edu}

\author{Jiaxin Mao}
\affiliation{%
	\institution{Gaoling School of Artificial Intelligence, Renmin University of China}
	\city{Beijing} 
	\country{China} 
}
\email{maojiaxin@gmail.com}

\begin{abstract}

\iftrue
How to obtain an unbiased ranking model by learning to rank with biased user feedback is an important research question for IR.
Existing work on unbiased learning to rank (ULTR) can be broadly categorized into two groups -- the studies on unbiased learning algorithms with logged data, namely the \textit{offline} unbiased learning, and the studies on unbiased parameters estimation with real-time user interactions, namely the \textit{online} learning to rank.
While their definitions of \textit{unbiasness} are different, these two types of ULTR algorithms share the same goal -- to find the best models that rank documents based on their intrinsic relevance or utility.
However, most studies on offline and online unbiased learning to rank are carried in parallel without detailed comparisons on their background theories and empirical performance.
In this paper, we formalize the task of unbiased learning to rank and show that existing algorithms for offline unbiased learning and online learning to rank are just the two sides of the same coin.
We evaluate eight state-of-the-art ULTR algorithms and find that many of them can be used in both offline settings and online environments with or without minor modifications.
Further, we analyze how different offline and online learning paradigms would affect the theoretical foundation and empirical effectiveness of each algorithm on both synthetic and real search data.
Our findings provide important insights and guideline for choosing and deploying ULTR algorithms in practice. 
\fi

\end{abstract}

%
%



\maketitle

\renewcommand{\shortauthors}{Q. Ai et al.}


\section{Introduction}

The study of learning to rank with implicit user feedback such as click data has received considerable attention in both academia and industry~\cite{joachims2005accurately}.
On the one hand, by collecting data from user interactions, we can better capture the true utility of each document for each user~\cite{wang2016learning} and create large-scale training data for ranking optimization without extensive human annotations~\cite{joachims2007evaluating}.
On the other hand, learning to rank directly with implicit user feedback could suffer from the intrinsic noise and bias in user interactions (e.g., position bias~\cite{craswell2008experimental}).
How to learn an unbiased learning to rank models from biased user feedback is thus an important question for the IR community.






Existing research on unbiased learning to rank (ULTR) algorithms can be broadly categorized into two groups. 
The first group focuses on creating robust learning algorithms that protect ranking models from inheriting data bias from observed data in the training process~\cite{joachims2017unbiased,ai2018unbiased,wang2018position,hu2019unbiased}. 
Because they can work with search logs or historical data, these types of algorithms are often referred to as the \textit{offline learning} methods, or, in most literature, the standard unbiased learning to rank algorithms.
The second group focuses on designing an interactive learning process so that we can collect unbiased feedback or estimate unbiased gradients for the training process of ranking models~\cite{yue2009interactively,oosterhuis2018differentiable,wang2018efficient}.
These types of methods require real-time interactions with end users on ranking results in each learning step, so they are often referred to as the \textit{online learning} algorithms, or, in most literature, the online learning to rank algorithms.


While both groups of methods have been widely studied under the topic of unbiased learning to rank, their definitions of ``unbiasness'' are slightly different. 
In the studies of standard unbiased learning-to-rank algorithms (with offline data), unbiasness usually refers to the ability of an algorithm in terms of removing the effect of data bias in the training of a ranking model~\cite{joachims2016counterfactual}.
The concept of unbiasness in online learning to rank, on the other hand, emphasizes more on whether an algorithm can help a model converge to the best ranking model for a particular task~\cite{wang2019variance}.
Despite these differences, both offline unbiased learning-to-rank algorithms and online learning to rank share a single goal for ranking optimization, that is \textit{to find the best model that ranks query-document pairs according to their intrinsic relevance.}

Then the question is: are offline learning and the online learning just two sides of the same coin for unbiased learning to rank?
The answer looks to be ``no'' considering their different motivations and definitions.
However, after examining the formulations of several algorithms, we observe that almost all unbiased learning-to-rank algorithms in offline learning can be directly applied to online learning, and some methods in online learning can be directly used on offline data without or with minor modifications.  
Does this mean that a good unbiased learning to rank algorithm could be used in both offline settings and online settings?
Or are there any properties that make an algorithm only suitable for offline learning or online learning?
Unfortunately, most research on offline unbiased learning to rank and online learning to rank are carried in parallel without comparisons.
Some recent studies tried to compare the empirical performance of existing unbiased learning-to-rank algorithms proposed separately in offline learning and online learning environments~\cite{Jagerman:2019:MIC:3331184.3331269}, but they are restricted by the stereotype that ``offline'' methods can only be used offline and ignores the discussion of the connections between offline unbiased learning and online learning to rank in theory.

In this paper, we conduct a comprehensive analysis on eight state-of-the-art unbiased learning to rank algorithms from two families -- the counterfactual learning family and the bandit learning family -- and discuss their characteristics in both offline and online learning. 
Specifically, we focus on two research questions:

\vspace{5pt}
\textbf{RQ1}: \textit{What are the theoretical differences and connections between unbiased learning-to-rank algorithms proposed for offline learning and online learning?} 

\vspace{5pt}
\textbf{RQ2}: \textit{How do learning paradigms affect the empirical effectiveness and robustness of unbiased learning-to-rank algorithms?}
\vspace{5pt}

\noindent 
To answer these questions, we develop a unified mathematical framework for unbiased learning to rank and discover that previous unbiased learning-to-rank algorithms proposed for offline and online learning are tackling the same problem from two perspectives.
Also, from our empirical studies with synthetic and real click data, we find that different unbiased learning-to-rank algorithms have different sensitivities to learning paradigms.  
For example, the performance of counterfactual learning algorithms are stable in both offline and online learning paradigms, while the effectiveness and robustness of bandit learning algorithms vary significantly from case to case. 
These findings provide both theoretical insights to the problem of unbiased learning to rank and practical guidelines to the deployment of unbiased learning-to-rank algorithms. 

The rest of the paper is organized as the followings.
Section~\ref{sec:related} discuss the related work of this paper.
Section~\ref{sec:problem} formally formulates the problem of unbiased learning to rank.
Then, Section~\ref{sec:theory} introduces the theoretical background of existing unbiased learning-to-rank algorithms.
Section~\ref{sec:deployment} discusses how to deploy each unbiased learning-to-rank algorithm in different learning paradigms and Section~\ref{sec:exp} shows our empirical experiments.
Finally, we summarize our findings to provide guidelance for the future use of unbiased learning-to-rank algorithms in Section~\ref{sec:conclusion}.


\section{Related Work}\label{sec:related}
Learning to rank refers to the machine learning techniques for training a ranking model~\cite{li2011short}.
It has been successfully and widely applied to multiple IR-related areas and applications such as ad-hoc retrieval~\cite{liu2009learning}, Web search~\cite{joachims2002optimizing}, question answering~\cite{yang2016beyond}, recommendation~\cite{duan2010empirical}, etc.
In general, given a set of feature representations for candidate items, the goal of learning to rank is to build a ranking function or scoring function that takes the features of an item as inputs and predicts a ranking score for it so that sorting items with their ranking scores could maximize the system's or the user's information gain from the final ranking list.
Based on their structures and definitions, there are two methodologies to classify existing learning-to-rank models.
The first one, which is also the most well-known one, is to categorize learning-to-rank models according to their training loss functions.
According to how many items are considered in the computation of ranking loss for each training instance~\cite{liu2009learning}, learning to rank algorithms can be categorized as pointwise, pairwise, or listwise approaches. 
Pointwise methods treat a ranking problem as a classification or regression problem by directly requiring the scoring function to predict the relevance label of the document~\cite{li2008mcrank}.
Pairwise methods transform ranking problems into a set of pairwise preference prediction tasks and compute ranking losses by aggregating errors of preferences indicated by ranking scores for each document pair~\cite{burges2005learning, joachims2002optimizing}.
Further, listwise methods extend pairwise methods by taking a set of documents together and directly optimizing the final ranking metrics~\cite{cao2007learning, burges2010ranknet, ai2018DLCM}.
Another classification methodology for learning-to-rank models is to categorize them based on the structure of ranking or scoring functions.
According to the number of input items the scoring function takes in each step, learning-to-rank algorithms can be grouped as univariate methods and multivariate methods.
Univariate methods assume that the relevance of each document are independent to each other, so the scoring function of a learning-to-rank model only needs to score one document a time~\cite{liu2009learning}.
Multivariate methods believe that the ranking and utility of documents could vary according to their context (i.e., other documents to rank) and design scoring functions by taking and comparing multiple documents together to determine their final ranking scores~\cite{ai2018DLCM,ai2019learning,pang2019setrank,pasumarthi2019self,yang2020analysis}.

While proven effective in improving the ranking performance~\cite{chapelle2011yahoo}, training learning-to-rank models usually requires large-scale data with annotated relevance labels that are expensive and time-consuming to collect. 
To solve this problem, IR researchers have tried to leverage the implicit feedback from user behavior as an alternative data source for training ranking models~\cite{joachims2002optimizing}. 
However, implicit feedback such as user clicks is noisy and affected by different kinds of biases~\cite{joachims2007evaluating, keane2006modeling, mao2018constructing, wang2013incorporating}, e.g., the ranking position has a strong influence on where users click~\cite{joachims2005accurately}.
This position bias makes the click signal a biased estimation of relevance and affects the robustness of learning to rank models trained with the na\"ive approach which treats a click/non-click as a positive/negative relevance judgment.
Thus, many studies have investigated how to extract unbiased and reliable relevance signals from biased click signals.
For example, \citet{joachims2002optimizing} proposed to treat clicks as preferences between clicked and skipped documents; \citet{richardson2007predicting} formalize an \emph{Examination Hypothesis} to model the position bias in the ranked list of ads by assuming that a user would only click a document when it is observed by the user and considered relevant to the user's need.
Accordingly, a series of click models  have been proposed to model the examination probability and infer accurate relevance feedback from user clicks~\cite{chapelle2009expected, mao2018constructing, wang2013incorporating, chuklin2015click}. 
\citet{craswell2008experimental} propose a cascade model to model user's sequential reading behavior on search engine result pages (SERPs).
\citet{dupret2008user} propose a user browsing model that allows users to read with jumps from previous results to latter results.
Nonetheless, despite their differences, click models usually requires that the same query-document pair appears multiple times for reliable inference~\cite{mao2019investigating}, which makes them invalid for tail queries and many retrieval tasks with special characteristics (e.g. email search).



Another group of approaches, which is the focus of this study, tried to directly train an unbiased ranking model with biased user feedback.
We refer to these approaches as the unbiased learning to rank (ULTR) approaches.
As mentioned earlier, existing unbiased learning-to-rank methods can be broadly classified into two families: the counterfactual learning family that originally adopts an offline learning paradigm and the bandit learning family that usually associate with an online learning paradigm.
The key of counterfactual learning algorithms is the Inverse Propensity Weighting (IPW)~\cite{wang2016learning,joachims2017unbiased} and the estimation of examination propensity~\cite{wang2018position,ai2018unbiased}.
For example, \citet{wang2016learning} propose an online result randomization experiments to estimate user's probabilities to examine result on each position and use the estimated weights to debias the training loss of learning-to-rank models.
\citet{ai2018unbiased} formulate the problem of ranking and examination propensity estimation as dual problems and build a Dual Learning Algorithm to automatically learn both ranking models and propensity models from offline data together.
There are also counterfactual learning algorithms that derive propensity estimation from online interleaving~\cite{joachims2017unbiased} or intervention data harvested from multiple ranking functions~\cite{agarwal2019estimating}.
Besides the counterfactual learning studies on position bias, there are studies on applying inverse propensity weighting to different behavioral biases such as the trust bias~\cite{agarwal2019addressing} and the recency bias~\cite{chen2019correcting}.

The core of the bandit learning algorithms is the estimation of unbiased model gradients from online result manipulation and user feedback.
A well-know example is the Dual Bandit Gradient Descent model proposed by \citet{yue2009interactively} that iteratively optimizes ranking models by creating random parameter perturbations and updating models with perturbed parameters that produces ranking lists with more user interactions.
There is extensive research on extending DBGD with different result exploration strategies~\cite{schuth2016multileave,zhao2016constructing,zhao2016constructing,wang2018efficient} and variance reduction techniques~\cite{wang2019variance}.
For example, \citet{schuth2014multileaved} propose to try multiple perturbed parameters simultaneously to speed up the convergence of bandit learning algorithms.
\citet{wang2018efficient} propose to store the noisy parameters explored previously and use null space analysis to find more efficient exploration directions in the future steps.
\citet{oosterhuis2018differentiable} propose to avoid extensive interleaving experiments by controlling the sample process of ranking lists and their weights in the training process. 
There are also algorithms developed independently with DBGD that combine click models~\cite{craswell2008experimental, guo2009efficient} with online bandit learning~\cite{kveton2015cascading, katariya2016dcm, zoghi2017online,lattimore2018toprank,li2018online}.
However, these click-model based methods usually estimate the utility of ranked documents on a per-query basis, which make them converge slower and less practical than other online algorithms such as PDGD (in Section~\ref{sec:bandit}).

While separate studies and tutorials on offline unbiased learning to rank~\cite{ai2018tutorial} and online unbiased learning to rank~\cite{grotov2016online} have been presented recently, to the best of our knowledge, there is no comprehensive analysis on their theoretical connections and differences.
\citet{Jagerman:2019:MIC:3331184.3331269} tried to compare counterfactual learning algorithms and bandit learning algorithms empirically, but they are confined by the stereotype that the former must be used offline and thus provided limited insights in theory.
Our study in this paper is timely and important for the understanding and applications of unbiased learning to rank in practice.

\section{Problem Definitions}\label{sec:problem}

\begin{table}[t]
	\caption{A summary of notations used in this paper.}
	\begin{tabular}
		{| p{0.09\textwidth} | p{0.8\textwidth}|} \hline
		$q$, $d$, $\pi_q$ & A query $q$ and a ranked list ($\pi_q$) of documents $d$ for $q$.   \\\hline
		$f$, $\theta$, $\phi$ & A ranking function $f$ parameterized by $\theta$ and an examination propensity model $\phi$. \\\hline
		$\mathcal{L}$, $l$, $\Delta$ & The global loss function ($\mathcal{L}$) of a ranking model, its local loss ($l$) on a query, and the loss ($\Delta$) on a specific document in the query.\\\hline
		$\mathbf{o}$, $\mathbf{r}$, $\mathbf{c}$ & Bernoulli variables that represent whether a document is observed ($o$), perceived as relevant ($r$) and clicked ($c$).\\\hline
	\end{tabular}\label{tab:notation}
\end{table}

In this section, we mathematically formalize the problem of unbiased learning to rank.
A summary of the notations used in this paper is shown in Table~\ref{tab:notation}. 

As discussed in previous studies~\cite{liu2009learning}, the goal of learning to rank is to learn a ranking function $f_{\theta}$ which takes the feature vector of document $d$ as input and produces a ranking score $f_{\theta}(d)$ so that ranking documents by $f_{\theta}(d)$ would result in the same ranked list as ranking documents by their intrinsic relevance, which we refer to as $\bm{r}$, to the query.
Formally, let $\mathcal{L}$ be a loss function of $f_{\theta}$, then learning to rank is to find the best $\theta$ (i.e., $\theta^*$) that 
\begin{equation}
\theta^* = \argmin_{\theta} \mathcal{L}(\theta) = \argmin_{\theta} \int_{q} l(f_{\theta}, \bm{r}_q)~dP(q)
\label{equ:ltr_goal}
\end{equation}
where $l(f_{\theta}, \bm{r}_q)$ is the local ranking loss computed based on the ranked list of documents and their relevance in each query session. 
The design of $l(f_{\theta}, \bm{r}_q)$ could vary according to the diverse needs of ranking tasks.
In general, most learning-to-rank algorithms would define $l(f_{\theta}, \bm{r}_q)$ according to positions of relevant and irrelevant documents based on similar design rationales used by ranking metrics such as MAP and NDCG~\cite{jarvelin2002cumulated}.


In noisy feedback environments such as Web search~\cite{joachims2005accurately} and e-commerce search~\cite{palotti2016learning}, relevance labels $\bm{r}$ are often inaccessible or unavailable. 
Instead, noisy user feedback that correlates to result relevance, e.g. user clicks, can be easily collected in large scale.
Similarly, we could conduct learning to rank with noisy user feedback and get the optimal ranking model with parameter $\theta^{\dagger}$ as
\begin{equation}
\theta^{\dagger} = \argmin_{\theta} \mathcal{L'}(\theta) = \argmin_{\theta} \int_{q} \! \int_{\pi_q} \mathbb{E}_{\bm{c}_{\pi_q}}[l'(f_{\theta}, \! \bm{c}_{\pi_q})]dP(q,\pi_q)
\label{equ:ultr_goal}
\end{equation}
where $\pi_q$ is the ranked list displayed in the session of $q$, $\bm{c}$ is the noisy feedback signals collected from users (e.g., clicks), and $P(q,\pi_q)$ is the probability of showing $\pi_q$ when a user submits a query $q$. 
We then can define the task of unbiased learning to rank as
\begin{definition}
	Given a loss function $\mathcal{L}$ computed based on the true relevance information $\bm{r}$, unbiased learning to rank is to find a function $\mathcal{L}'$ computed based on noisy labels $\bm{c}$ so that, for the optimal model parameters learned by $\mathcal{L}$ and $\mathcal{L}'$ (i.e., $\theta^*$ and $\theta^{\dagger}$, respectively), we have $\mathcal{L}(\theta^{\dagger}) = \mathcal{L}(\theta^*)$.
	\label{def:ultr}
\end{definition}
\noindent

\revised{Given this definition, we can see that the task of unbiased learning to rank involves two parts. 
The first part is ranking optimization. 
Given a particular loss function (e.g., $\mathcal{L}(\theta)$ or $\mathcal{L}'(\theta)$), whether Eq.~(\ref{equ:ltr_goal}) and Eq.~(\ref{equ:ultr_goal}) can be achieved through optimizations will directly determine whether it is possible to conduct unbiased learning to rank. 
In fact, the problem of ranking optimization is a general learning-to-rank problem that is not specific for unbiased learning to rank, and, in order to find the global optimum of $\theta$ in theory, different algorithms often have their unique requirements for ranking functions and loss functions.
For simplicity, we assume that all ranking functions and loss functions used in our theoretical analysis satisfy the requirements of each learning algorithm so that ranking optimization would not be an issue.
}

\revised{
The second part, which is the core of unbiased learning-to-rank algorithms, is whether $\mathcal{L}(\theta^{\dagger}) = \mathcal{L}(\theta^*)$ can be satisfied given Eq.~(\ref{equ:ltr_goal}) and Eq.~(\ref{equ:ultr_goal}).
Let the \textit{unbiasness} of a learning algorithm be how similar $\mathcal{L}(\theta^{\dagger})$ and $\mathcal{L}(\theta^*)$ would be.
}
Then, mathematically, we can derive two theorems for the definition of unbiased loss functions in unbiased learning to rank as:
\begin{theorem}
	$\mathcal{L'}(\theta)$ is unbiased if, for any $\theta$, $\mathcal{L'}(\theta)  = \mathcal{L}(\theta)$.
	\label{def:unbiased_counterfactual}
\end{theorem}
\noindent
and
\begin{theorem}
	$\mathcal{L}'(\theta)$ is unbiased if, for any $(\theta, \theta')$, 
	$\mathcal{L'}(\theta)\geq \mathcal{L'}(\theta')\Rightarrow \mathcal{L}(\theta)\geq \mathcal{L}(\theta')$.
	\label{def:unbiased_bandit}
\end{theorem}
\noindent 
The intuitive explanation of these theorems is that a loss function $\mathcal{L'}$ computed with click is unbiased if it is always equal to $\mathcal{L}$ or always has same preferences over model parameters with $\mathcal{L}$.
In this paper, we conduct analysis on eight representative and state-of-the-art unbiased learning to rank algorithms in both offline and online settings from the perspective of theoretical foundations and the perspective of practical deployment.

   




\section{Theoretical Foundations}\label{sec:theory}

In this section, we discuss the theory behind existing unbiased learning to rank algorithms and how they achieve or approximate \textit{unbiasness} as defined in Section~\ref{sec:problem}.
For simplicity, we focus on a standard ranking scenario where we retrieve and display documents sequentially in a list according to user's need (e.g., the ten blue links provided by Web search engines). 
Let $\bm{r}$ be the true relevance of each documents, $\bm{o}$ be the variable indicating whether the user has examined the documents, and $\bm{c}$ be the click behavior of search users.
According to well-established user examination hypothesis~\cite{richardson2007predicting}, each document will be click (i.e., $c_d=1$) if and only if the user has examined the document (i.e., $o_d=1$) and the document is relevant (i.e., $r_d=1$)~\cite{richardson2007predicting}.
In other words, we have 
\begin{equation}
P(c_d=1) = P(o_d=1)\cdot P(r_d=1)
\label{equ:click_assumption}
\end{equation}
\revised{Note that we assume that $P(o_d=1)>0$ here because no algorithms trained by click data can be unbiased in theory if there are result positions that users never examine.}
Also, as most ranking metrics (e.g. NDCG~\cite{jarvelin2002cumulated}, ERR~\cite{chapelle2009expected}) only concern about the position of relevant documents, we assume that the local loss function in Eq.~(\ref{equ:ltr_goal}) can be formulated as 
\begin{equation}
\begin{split}
l(f_{\theta}, \bm{r}) = \sum_{d, r_d=1}  \Delta(d, r_d|f_{\theta})
\end{split}
\label{equ:rank_loss}
\end{equation}
where $\Delta(d, r_d|f_{\theta})$ is a function that computes the individual loss on each relevant document for a ranking model $f_{\theta}$. 

Broadly speaking, existing unbiased learning to rank algorithms can be categorized into two groups.
The first group focuses on achieving unbiasness through Theorem~\ref{def:unbiased_counterfactual} by directly removing the inherited bias from user feedback in the computation of ranking loss, while the second group focuses on achieving unbiasness through Theorem~\ref{def:unbiased_bandit} by manipulating the process of user feedback collection to estimate unbiased gradients for model optimization.
Formally, in Eq.~(\ref{equ:ultr_goal}), the first group tries to solve unbiased learning to rank by developing new ranking loss function $l'(f_{\theta}, \bm{c}_{\pi_q})$, while the second group aims to tackle the problem by controlling the distribution of $P(q,\pi_q)$. 
Based on different motivations, we have two families of unbiased learning to rank algorithms designed under those two groups -- the counterfactual learning family and the bandit learning family.




\subsection{Counterfactual Learning Family}\label{sec:counterfactual}

The idea of counterfactual learning is to remove the effect of data bias in the computation of ranking loss $l'\!(f_{\theta}, \bm{c}_{\pi_q})$ so that the model trained with biased data (i.e., clicks) would converge to the model trained unbiased labels (i.e., the relevance of a document). 
Specifically, optimization in counterfactual learning can be simplified as
\begin{equation}
\mathcal{L'}(\theta) \!=\! \int_{q} \!\int_{\pi_q} \mathbb{E}_{\bm{c}_{\pi_q}}\![l'\!(f_{\theta}, \bm{c}_{\pi_q})]dP(q,\pi_q) \!=\! \int_{q}\! \mathbb{E}_{\bm{c}}\![l'\!(f_{\theta}, \bm{c})]dP(q) 
\label{equ:cl_goal}
\end{equation}
where we use $\bm{c}$ to represent the clicks observed in each search session.
Because it doesn't concern about the distribution of how results are displayed in each query session (i.e., $P(q,\pi_q)$), counterfactual learning naturally suits the need of learning to rank with historical data where the distribution of the logging systems may not be available.
Indeed, most studies on counterfactual unbiased learning to rank focus on the use of search logs and offline learning~\cite{wang2016learning,joachims2017unbiased,wang2018position,ai2018unbiased,agarwal2019general}.

Based on Eq.~(\ref{equ:cl_goal}), we introduce four algorithms in the rest of this section, which are the Inverse Propensity Weighting model~\cite{wang2016learning,joachims2017unbiased}, the Regression-based EM model ~\cite{wang2018position}, the Dual Learning Algorithm~\cite{ai2018unbiased}, and the Pairwise Debiasing model~\cite{hu2019unbiased}.

\subsubsection{Inverse Propensity Weighting}\label{sec:IPW}
Inverse Propensity Weighting (IPW) is one of the first ULTR algorithms proposed under the framework of counterfactual learning~\cite{wang2016learning,joachims2017unbiased}.
The basic idea of IPW is to revise the computation of $l(f_{\theta}, \bm{r})$ with user feedback data $\bm{c}$ as
\begin{equation}
\begin{split}
l'(f_{\theta}, \bm{c}) = l_{IPW}(f_{\theta}, \bm{c}) & = \sum_{\substack{d, c_d=1}} \frac{\Delta(d, c_d|f_{\theta})}{P(o_d =1)}
\end{split}
\label{equ:rank_IPW}
\end{equation}
where $P(o_d =1)$ is the probability of document $d$ being examined in the search session.
Joachims et al.~\cite{joachims2017unbiased} proves that the expectation of the $l_{IPW}(f_{\theta}, \bm{c})$ is equal to $l(f_{\theta}, \bm{r})$ as
\begin{equation}
\begin{split}
\mathbb{E}_{\mathbf{o}}[l_{IPW}(f_{\theta}, \bm{c})] &= \mathbb{E}_{\mathbf{o}}\big[\sum_{\substack{d, c_d=1}} \frac{\Delta(d, c_d|f_{\theta})}{P(o_d =1)}\big] \\
&= \mathbb{E}_{\mathbf{o}}\big[\sum_{d,r_d=1} \frac{o_d \cdot \Delta(d, r_d|f_{\theta})}{P(o_d =1)}\big] \\
&= \sum_{d, r_d=1} \mathbb{E}_{\mathbf{o}}\big[o_d\big] \cdot \frac{\Delta(d, r_d|f_{\theta})}{P(o_d =1)} \\
&= \sum_{d, r_d=1} P(o_d =1) \cdot \frac{\Delta(d, r_d|f_{\theta})}{P(o_d =1)} \\
&= \sum_{d, r_d=1} \Delta(d, r_d|f_{\theta}) \\
&= l(f_{\theta}, \bm{r})
\end{split}
\label{equ:ipw_expectation}
\end{equation}
Thus, according to Theorem~\ref{def:unbiased_counterfactual}, IPW is theoretically principled for unbiased learning to rank.

The key of IPW is the estimation of examination propensity (i.e., $P(o_d=1)$).
Assuming that the examination of documents only depends on their positions in ranked lists, Wang et al.~\cite{wang2016learning} and Joachims et al.~\cite{joachims2017unbiased} conducted online result randomization to estimate $P(o_d =1)$.
The idea of result randomization is to randomly shuffle the documents in each query so that relevant documents would have equal probabilities to be placed at each position in ranked lists. 
Let $c_k$ be the clicks on the $k$th position of a ranked list, then we have $P(r_k=1)$ equivalent to a constant and
\begin{equation}
\begin{split}
\mathbb{E}[c_k] &= \int_{(q,\pi_q)}\!\!\!\!\!\!c_k~dP(q,\pi_q) \\
&= \int_{(q,\pi_q)}\!\!\!\!\!\!o_k \cdot r_k ~ dP(q,x,\pi_q) \\
&= P(o_k=1) \cdot P(r_k=1) \\
&\propto P(o_k=1) 
\end{split}
\label{equ:random}
\end{equation}
where $o_k$ is whether the user has examined the document on the $k$th position, and we use the fact that $o_k$ and $r_k$ are independent in the third step of Eq.~(\ref{equ:random}). 
Therefore, we could directly estimate $P(o_k=1)$ using user clicks (i.e., $\mathbb{E}[c_k]$) with online result randomization.
Though harmful to user experience, this method creates an unbiased estimation of $P(o_d =1)$ in theory~\cite{wang2018position,ai2018unbiased}, which consequentially guarantees the unbiasness of IPW.



\subsubsection{Regression EM}
To avoid hurting user experience with online result randomization, Wang et al.~\cite{wang2018position} propose to unify the training of ranking models and the estimation of examination propensity with a graphic model and an EM algorithm, which we refer to as the Regression EM Model (REM).
Based on the user examination hypothesis described in Eq.~(\ref{equ:click_assumption}), REM computes the likelihood of observed clicks for each query $q$ as 
\begin{equation}
\begin{split}
\log P(\bm{c}) = \sum_{d} c_d\log(P(o_d=1)\cdot P(r_d=1)) + (1-c_d)\log(1-P(o_d=1)\cdot P(r_d=1))
\end{split}
\label{equ:REM}
\end{equation}
where $c_d$ can be observed from search logs or online user interactions, and $o_d$ and $r_d$ are latent variables. 
Specifically, $P(r_d=1)$ is computed based on the ranking function $f_{\theta}$ as 
$$
P(r_d=1) = \frac{1}{1+\exp(-f_{\theta}(d))}
$$
Using standard EM algorithms, we can estimate $o_d$ and $f_{\theta}(d)$ based on observed click logs and Eq.~(\ref{equ:REM}).
The estimation is guaranteed by EM algorithms~\cite{cappe2009line} to be unbiased for the pointwise loss function $l$ like
$$
l(f_{\theta}, \bm{r}) = -\sum_{d}r_d\log(P(r_d=1)) + (1-r_d)\log(P(r_d=1))   
$$ 
Therefore, REM is a theoretically principled ULTR algorithm under Theorem~\ref{def:unbiased_counterfactual}. 


\subsubsection{Dual Learning Algorithm}
Proposed with REM in parallel, Dual Learning Algorithm (DLA)~\cite{ai2018unbiased} also tries to conduct unbiased learning to rank without result randomization. 
The key observation in DLA is that the positions of $o_d$ and $r_d$ are interchangeable in Eq.~(\ref{equ:click_assumption}), which means that, in theory, the counterfactual learning of inverse propensity weighting can be applied on both directions simultaneously. 
Specifically, Ai et al.~\cite{ai2018unbiased} propose to treat the estimation of examination propensity as a dual problem of learning to rank and learn a propensity model $\bm{\phi}$ by optimizing an inversed relevance weighted loss function (IRW) as
\begin{equation}
	l_{IRW}(\bm{\phi}, \bm{c})  = \sum_{\substack{d, c_d=1}} \frac{\Delta(d, c_d|\bm{\phi})}{P(r_d =1)}
\end{equation}
Suppose that the expectation of $P(\bm{r})$ on each position (over all sessions) is stable, then we have similar induction to Eq.~(\ref{equ:ipw_expectation}) as
\begin{equation}
\begin{split}
\mathbb{E}_{\mathbf{r}}[l_{IRW}(\bm{\phi}, \bm{c})] &= \mathbb{E}_{\mathbf{r}}\big[\sum_{\substack{d, c_d=1}} \frac{\Delta(d, c_d|\bm{\phi})}{P(r_d =1)}\big] \\
&= \mathbb{E}_{\mathbf{r}}\big[\sum_{d,o_d=1} \frac{r_d \cdot \Delta(d, o_d|\bm{\phi})}{P(r_d =1)}\big] \\
&= \sum_{d, o_d=1} \mathbb{E}_{\mathbf{r}}\big[r_d\big] \cdot \frac{\Delta(d, o_d|\bm{\phi})}{P(r_d =1)} \\
&= \sum_{d, o_d=1} P(r_d =1) \cdot \frac{\Delta(d, o_d|\bm{\phi})}{P(r_d =1)} \\
&= \sum_{d, o_d=1} \Delta(d, o_d|\bm{\phi}) \\
&= l(\bm{\phi}, \bm{o})
\end{split}
\label{equ:irw_expectation}
\end{equation}
which means that IRW is an unbiased estimation of examination propensity model $\bm{\phi}$.
Thus, if we model $P(r_d=1)$ and $P(o_d=1)$ with $f_{\theta}(d)$ and $\phi_d$, DLA can iteratively conduct unbiased learning for both $P(r_d=1)$ and $P(o_d=1)$ with inverse propensity weighting and inverse relevance weighting.
As a better ranking model $f_{\theta}(d)$ will lead to a better estimation of the propensity model $\phi_d$ and a better $\phi_d$ will also lead to a better estimation of $f_{\theta}(d)$, DLA is guaranteed to converge to the unbiased ranking model and unbiased propensity model simultaneously in theory~\cite{ai2018unbiased}.

\subsubsection{Pairwise Debiasing}
The Pairwise Debiasing (PairD) model is a variation of DLA with pairwise ranking losses proposed by Hu et al.~\cite{hu2019unbiased}.
Specifically, it conducts learning to rank with inverse propensity weighting and estimates examination propensity models together with the ranking models. 
The differences between DLA and PairD are two-fold. 
First, PairD is specifically designed for pairwise learning to rank where $l(f_{\theta}, \bm{r})$ can be formulated as a sum of pairwise losses $\Delta(f_{\theta}, d^+, d^-)$ where $r_{d^+} > r_{d^-}$.
Second, PairD tries to consider unclicked documents in its learning process by assuming that~\cite{hu2019unbiased}
\begin{equation}
\begin{split}
P(c_d=0) = t\cdot P(r_d=0)
\end{split}
\label{equ:pd_assumption}
\end{equation}
and computes the inverse propensity weighted version of $l'(f_{\theta}, \bm{c})$ as
\begin{equation}
\begin{split}
l'(f_{\theta}, \bm{c}) = l_{PairD}(f_{\theta}, \bm{c}) & = \sum_{\substack{d^+, d^-, c_{d^+}=1, c_{d^-}=0}} \frac{\Delta(f_{\theta}, d^+, d^-)}{P(o_d =1)\cdot t}
\end{split}
\label{equ:pd_ipw}
\end{equation}
Similar to REM, PairD adopts a EM-style greedy algorithm to estimate $P(o_d=1)$ and $t$ according to the observed clicks and current ranking model $f_{\theta}$.
For simplicity, we ignore the derivations of optimization loss for $P(o_d =1)$ and $t$ in this paper.

Unfortunately, PairD is not theoretically unbiased because its assumption in Eq.~(\ref{equ:pd_assumption}) contradicts to the basic examination hypothesis that a document can only be clicked when it is examined and considered relevant by the user.
According to Eq.~(\ref{equ:click_assumption}), we have the probability of a document being not clicked as 
\begin{equation}
P(c_d=0) = 1 - P(r_d=1)\cdot P(o_d=1) = 1-P(o_d=1) +  P(o_d=1) \cdot P(r_d=0)
\label{equ:standard_examiniation}
\end{equation}
\revised{As we can see, the click assumption of PairD in Eq.~(\ref{equ:pd_assumption}) is different from the standard examination hypothesis depicted in Eq.~(\ref{equ:standard_examiniation}) because of a constant (i.e., $1-P(o_d=1)$) in Eq.~(\ref{equ:standard_examiniation}) that varies according to $P(o_d=1)$.
This constant is not neglectable given the fact that both $P(o_d=1)\in [0,1]$ and $P(r_d=0)\in [0,1]$. 
Therefore, the estimation of ranking models in PairD is consequentially not unbiased under any analysis based on the user examination hypothesis~\cite{richardson2007predicting}.	
While its theoretical foundation is controversial, PairD's empirical performance is not bad as Eq.~(\ref{equ:pd_assumption}) could be a reasonable approximation of Eq.~(\ref{equ:standard_examiniation}) in some cases (e.g., when $P(o_d=1)$ is large), which means that PairD can alleviate click bias in some degrees in its pairwise loss and counterfactual process. 
}

\subsection{Bandit Learning Family}\label{sec:bandit}

Bandit learning aims to collect real-time user feedback in controlled environments so that we can explain and analyze the observed data and update models accordingly.
In unbiased learning to rank, this means estimating unbiased parameter gradients from click data by controlling the displayed ranked lists for each query in each session.
Mathematically, ranking optimization in bandit learning can be reformulated as  
\begin{equation}
\mathcal{L'}(\theta) = \int_{q} \int_{\pi_q} \mathbb{E}_{\bm{c}_{\pi_q}}[l(f_{\theta}, \bm{c}_{\pi_q})]~dP(\pi_q|q)dP(q) 
\label{equ:bl_goal}
\end{equation}
where we simply replace $\bm{r}$ with $\bm{c}_{\pi_q}$ in Eq.~(\ref*{equ:ltr_goal}) and manipulate $P(\pi_q|q)$ to achieve unbiasness.
In other words, the goal of unbiased bandit learning to rank is to find $P(\pi_q|q)$ so that for any $(\theta, \theta')$:
\begin{equation}
\begin{split}
\int_{\pi_q} \mathbb{E}_{\bm{c}_{\pi_q}}\![l(f_{\theta}, \!\bm{c}_{\pi_q})]~dP(\pi_q|q) \geq &\int_{\pi_q} \mathbb{E}_{\bm{c}_{\pi_q}}\![l(f_{\theta'},\! \bm{c}_{\pi_q})]~dP(\pi_q|q) \\
\Rightarrow  l(f_{\theta}, \bm{r}) \geq & ~~l(f_{\theta'}, \bm{r})
\end{split}
\label{equ:bl_sub_goal}
\end{equation}
As bandit learning can reuse $l(f_{\theta}, \bm{r})$ on click data, algorithms under this family are easier to design and analyze~\cite{joachims2016counterfactual}. 
However, because they require the control of $P(\pi_q|q)$, bandit learning algorithms are often used in online environments, which is exactly the reason of why they are commonly referred to as the \textit{online learning to rank} algorithms.
Next, we briefly discuss a couple of classical and state-of-the-art algorithms in the bandit learning family, which are the Dueling Bandit Gradient Descent, Multileave Gradient Descent, Null Space Gradient Descent, and the Pairwise Differentiable Gradient Descent algorithm.

\subsubsection{Dueling Bandit Gradient Descent}\label{sec:DBGD}
Proposed by Yue and Joachims~\cite{yue2009interactively}, the Dueling Bandit Gradient Descent (DBGD) algorithm optimizes $f_{\theta}$ with three steps:
\begin{itemize}
	\item \textit{Step 1}: Sample (usually uniformly) a parameter perturbation and add it to the original parameter $\theta$ to form a perturbed parameter $\theta'$ so that $f_{\theta}$ and $f_{\theta'}$ produce different ranked lists $\pi_{\theta}$ and $\pi_{\theta'}$, respectively.
	\item \textit{Step 2}: Show $\pi_{\theta}$ and $\pi_{\theta'}$ (directly or interleavedly) to real users to collect clicks and compute $l(f_{\theta}, \bm{c}_{\pi_{\theta}})$ and $l(f_{\theta'}, \bm{c}_{\pi_{\theta'}})$.
	\item \textit{Step 3}: Upadte $\theta$ with $\theta'$ if we observe $\mathbb{E}_{\bm{c}_{\pi_{\theta'}}}[l(f_{\theta'}, \bm{c}_{\pi_{\theta'}})] < \mathbb{E}_{\bm{c}_{\pi_{\theta}}}[l(f_{\theta}, \bm{c}_{\pi_{\theta}})]$ in online experiments.
\end{itemize}
By repeating these steps, DBGD can gradually improve $f_{\theta}$ and achieve unbiased learning to rank in online environments.

The theoretical foundation of DBGD is established on the fact that, when both examined by users, relevant documents are more likely to be clicked than irrelevant documents (i.e., Eq.~(\ref{equ:click_assumption})). 
Assuming that user's behavior models would not change during the time period of online experiments, statistically, a ranking list produced by a better ranking model will receive more clicks. 
Therefore, \revised{assuming that the ranking functions and loss functions satisfy the requirements of DBGD in ranking optimization~\cite{yue2009interactively}}, the statistical expectation of $l(f_{\theta'}, \bm{c}_{\pi_{\theta'}})$ is smaller than $l(f_{\theta}, \bm{c}_{\pi_{\theta}})$ if and only if $l(f_{\theta'}, \bm{r}) < l(f_{\theta}, \bm{r})$, which makes DBGD an unbiased learning algorithm in theory according to Theorem~\ref{def:unbiased_bandit}.  


\subsubsection{Multileave Gradient Descent and Null Space Gradient Descent}
Multileave Gradient Descent (MGD)~\cite{schuth2016multileave} and Null Space Gradient Descent (NSGD)~\cite{wang2018efficient} are well-known extensions to the original DBGD algorithm.
Overall, the structures of MGD and NSGD are similar to DBGD -- they all share the same three-step optimization process as described in Section~\ref{sec:DBGD}.
The only differences between MGD, NSGD, and DBGD is how they sample the perturbed parameter $\theta'$ in \textit{Step 1}. 
Instead of sampling $\theta'$ once in each iteration, MGD samples multiple $\theta'$ every time and combine the ranked lists of all sampled $\theta'$ to collect clicks and update models.
It essentially explores more possible parameter settings than DBGD in each iteration.
NSGD follows a similar methodology with MGD but use a different sample strategy.
It stores the explored perturbed parameters in previous iterations and sample new parameters from the null space (i.e., the space of vectors that are orthogonal to the target vectors) of previous noisy parameters to avoid exploring directions that have been proven ineffective in previous training iteration.

Theoretically, there is no difference between DBGD, MGD, and NSGD in terms of unbiasness.
As they always choose parameters that produce ranking lists with more user clicks, they are guaranteed to be unbiased based on Theorem~\ref{def:unbiased_bandit}. 
In practice, however, whether each of them could produce effective learning-to-rank models highly depends on data characteristics, model structures, parameter initialization, and many other factors other than the learning algorithms themselves.  
In general, we observe great variance in the training of DBGD, MGD, and NSGD with ranking models such as deep neural networks.
\revised{Part of the reason is that, despite of their well-developed theoretical background, DBGD-based bandit learning algorithms often have strict restrictions and assumptions on the design of ranking functions and loss functions for ranking optimization.
For instance, DBGD assumes that the utility function (i.e., the loss function) is strictly concave given the ranking function $f_{\theta}$ so that the optimal value of $\theta$ can be unique (i.e., the Lipschitz smoothness assumption~\cite{yue2009interactively}).
As shown by \citet{oosterhuis2019optimizing}, this requirement is not easy to achieve in practice with deterministic linear models or other simple ranking functions.
}
We will discuss more of their empirical performance in Section~\ref{sec:exp}.

\subsubsection{Pairwise Differentiable Gradient Descent}\label{sec:PDGD}

The motivation of Pairwise Differentiable Gradient Descent (PDGD)~\cite{oosterhuis2018differentiable} is the observation that we can infer unbiased gradients from clicks on a single $\pi_q$ by controlling the prior distribution of $P(\pi_q|q)$. 
Let $\pi_q$ be stochastically sampled according to $f_{\theta}$ and the Plackett-Luce model as 
\begin{equation}
\begin{split}
P(\pi_q|q) = \prod_{i=1}^{|\pi_q|} \frac{\exp(f_{\theta}(d_i))}{\sum_{j=i}^{|\pi_q|}\exp(f_{\theta}(d_j))}
\label{equ:stochastic_ranking}
\end{split}
\end{equation}
where $d_i$ is the $i$th document in $\pi_q$.
Assuming that users always read $\pi_q$ sequentially~\cite{craswell2008experimental}, PDGD treats $l(f_{\theta}, \bm{c})$ as the sum of pairwise losses $\Delta(f_{\theta}, d_i, d_j)$ over document pairs as  
\begin{equation}
\begin{split}
\Delta'(f_{\theta}, \bm{c}_{\pi_q}) = \sum_{\substack{d_i, d_j, j < i+2, c_{d_i}=1, c_{d_j}=0}}  \rho(d_i, d_j, \pi_q) \cdot \Delta(f_{\theta}, d_i, d_j)
\label{equ:pdgd_loss}
\end{split}
\end{equation}
where $d_i$ is a clicked document behind $d_j$ in $\pi_q$, and $\rho(d_i, d_j, \pi_q)$ is 
\begin{equation}
\begin{split}
	\rho(d_i, d_j, \pi_q) = \frac{P(\pi_q(d_i, d_j)|q)}{P(\pi_q|q) + P(\pi_q(d_j, d_i)|q)}
	\label{equ:pdgd_weight}
\end{split}
\end{equation}
where $\pi_q(d_j, d_i)$ is the ranking of documents after reversing the position of $d_i$ and $d_j$ in $\pi_q$.

\revised{The proof of unbiasness for PDGD is complicated.
In the original paper of PDGD, \citet{oosterhuis2018differentiable} prove that PDGD is unbiased with respect to the pairwise loss on each pair of documents.
}
For example, suppose that there is a query with only 2 candidate documents, i.e. $d_a$ and $d_b$, which means that there are only two possible ranked lists, i.e. $\pi_{ab}=[d_a,d_b]$ and $\pi_{ba}=[d_b,d_a]$.
Let $P(c^{2nd}_{ab})$ be the probability of observing a click on the 2nd position in all sessions that show $\pi_{ab}$ to the user.
Because of the stochastic sampling process in Eq.~(\ref{equ:stochastic_ranking}), the probability of $d_a$ and $d_b$ being the 2nd document in $\pi_q$ and receiving clicks would be $P(c^{2nd}_{ab})P(\pi_{ab}|q)$ and $P(c^{2nd}_{ba})P(\pi_{ba}|q)$, respectively. 
Let $\rho_{ab}$ and $\Delta_{ab}$ be the short forms of $\rho(d_a, d_b, \pi(a,b))$ and $\Delta(f_{\theta}, d_a, d_b)$, then we have
\begin{equation}
\begin{split}
 l(f_{\theta}, \bm{c}^{2n\!d}\!) \!=&\int_{\pi_q} \Delta'(f_{\theta}, \bm{c}_{\pi_q}^{2nd})~dP(\pi_q|q) \\
 =& P(c^{2nd}_{ab})P(\pi_{ab}|q)\rho_{ba}\Delta_{ba} + P(c^{2nd}_{ba})P(\pi_{ba}|q)\rho_{ab}\Delta_{ab}  \\
 =& \frac{P(\pi_{ba}|q)P(\pi_{ab}|q)}{P(\pi_{ba}|q) + P(\pi_{ab}|q)} (P(c^{2nd}_{ab})\Delta_{ba} + P(c^{2nd}_{ba})\Delta_{ab}) \\
 =& \frac{P(\pi_{ba}|q)P(\pi_{ab}|q)}{P(\pi_{ba}|q) + P(\pi_{ab}|q)} (P(c^{2nd}_{ab}) - P(c^{2nd}_{ba}))\Delta_{ba} \\
 \propto & (P(r_b=1) - P(r_a=1))\Delta_{ba}
\label{equ:pdgd_example}
\end{split}
\end{equation}
where we use $\Delta_{ab} = - \Delta_{ba}$ (which holds for most pairwise loss) in the third step, and use $P(c^{2nd}_{ab})=P(o^{2nd}\!=1)P(r_b=1)$ in the fourth step.
\revised{Therefore, $l'(f_{\theta}, \bm{c}^{2nd})$ is proportional to the pairwise preferences of $d_a$ and $d_b$ based on their true relevance $r_a$ and $r_b$, which, according to Theorem~\ref{def:unbiased_bandit}, means that PDGD provides unbiased loss estimation for clicks on ranked lists with 2 documents.
}

\revised{
However, when considering ranked lists with more than 2 documents, whether the unbiasness of PDGD still holds remain to be unknown.
As discussed previously, PDGD optimizes ranking by minimizing the sum of pairwise losses over document pairs as defined in Eq.~(\ref{equ:pdgd_loss}).
While \citet{oosterhuis2018differentiable} has shown that the individual pairwise loss of each document pair in PDGD is proportional to their pairwise loss computed with relevance labels, this doesn't indicate that the sum of those individual pairwise losses would still be proportional to the sum of relevance-based pairwise losses.
Thus, in this paper, we can only conclude that PDGD is \textit{partially unbiased} given specific restrictions and leave the proof of complete unbiasness for PDGD in future studies.
}

\subsection{Discussion}

As discussed previously, existing algorithms in the counterfactual learning family and the bandit learning family tries to tackle the same problem from two different perspectives.
The former focuses on the design of $l'(f_{\theta}, \bm{c})$ to achieve Theorem~\ref{def:unbiased_counterfactual}, while the latter tries to manipulate $P(q,\pi_q)$ to achieve Theorem~\ref{def:unbiased_bandit}.  
From this perspective, the theories behind counterfactual learning and bandit learning are essentially the two sides of the same coin.
The main reason why the studies of unbiased learning to rank with counterfactual learning are often conducted separately from those with bandit learning (which are more often referred to as \textit{online learning to rank}) is that counterfactual learning can be applied on offline data while bandit learning can only be used in online environments.

In fact, the connection between counterfactual learning and bandit learning is stronger than it is appeared to be.
For example, to estimate examination propensity for IPW, Wang et al.~\cite{wang2016learning} and Joachims et al.~\cite{joachims2017unbiased} propose to conduct online result randomization, which randomly shuffle the positions of all documents before showing them to the users.
In this case, $P(\pi_q|q)$ is a uniform distribution over the universal set of possible ranked lists $\Omega_q$ (i.e., $P(\pi_q|q) = 1/|\Omega_q|$).
If we use the clicks on the $i$th position from online result randomization to compute pairwise loss on documents $d_a$ and $d_b$, then we have
\begin{equation}
\begin{split}
 l'\!(f_{\theta}, \bm{c}^{i}\!) =& \int_{\pi_a^i} \!P(c_a^i=1)\Delta_{ba} dP(\pi_a^i|q) \!+\! \int_{\pi_b^i} \!P(c_b^i=1)\Delta_{ab} dP(\pi_b^i|q) \\
=& P(c_a^i=1)P(\pi_{a}^i|q)\Delta_{ba} + P(c^{i}_{b}=1)P(\pi_{b}^i|q)\Delta_{ab}  \\
=& \frac{1}{|\Omega_q|} (P(c_a^i=1) - P(c_b^i=1))\Delta_{ba} \\
=& \frac{P(o_i=1)}{|\Omega_q|} (P(r_b=1) - P(r_a=1))\Delta_{ba} \\
\propto & (P(r_b=1) - P(r_a=1))\Delta_{ba}
\label{equ:random_example}
\end{split}
\end{equation}
where $\pi_a^i$ is the ranked list where $d_a$ is at the $i$th position, and $P(c_a^i=1)$ is the probability of $d_a$ being clicked when it is at the $i$th position.
Comparing Eq~(\ref{equ:pdgd_example}) and (\ref{equ:random_example}), we can see that 
\textit{bandit learning is essentially an online result randomization with controlled prior distributions of document ranking}.
In other words, if we use online randomization to estimate examination propensity for unbiased learning to rank, we could have the counterfactual learning algorithms; if we use online randomization to estimate relative document relevance directly, we could have bandit learning algorithms.



\section{Practical Deployments}\label{sec:deployment}


In this section, we discuss the deployments of existing unbiased learning to rank algorithms in practice. Specifically, we analyze how and why different offline and online learning paradigms would affect the effectiveness of algorithms in the counterfactual learning family and the bandit learning family.

\subsection{Offline or Online}

Although the research on unbiased learning to rank in offline settings and online settings has mostly been carried out in parallel, there is, to the best of our knowledge, no study that illustrates why offline ULTR methods cannot be used in online learning and why online ULTR algorithms are not applicable on offline data.
In fact, we observe that almost all algorithms in the counterfactual learning family can be applied in online learning environments, and some methods in the bandit learning family are applicable to offline data. 

In this paper, we focus our analysis on three types of learning paradigms.
The first one, which we refer to as the \textbf{offline paradigm} (\textit{Off}), is a classic setting where we train a ranking function $f_{\theta}$ based on the click logs collected from an existing system. In this paradigm, both the displayed ranked list ($\pi_q$) and the clicks on it ($\bm{c}_{\pi_q}$) are fixed and observed in advance.
The second one, which we refer to as the \textbf{stochastic online paradigm} (\textit{OnS}), is an online setting where $\pi_q$ is dynamically sampled with the Plackett-Luce model in Eq.~(\ref{equ:stochastic_ranking}) according to the current state of $f_{\theta}$, and $f_{\theta}$ is updated based on $\bm{c}_{\pi_q}$ collected online.
The third setting, which we refer to as the \textbf{deterministic online paradigm} (\textit{OnD}), is same to the stochastic online paradigm except that $\pi_q$ is created by ranking documents with $f_{\theta}$ directly.

Theoretically, algorithms in the counterfactual learning family can be applied in both offline and online settings because they have no requirement on $\pi_q$.
We could easily create three variations for each counterfactual learning to rank algorithm.
For example, we have IPW$_{Off}$, IPW$_{OnS}$, and IPW$_{OnD}$ for inverse propensity weighting algorithms with the offline learning, the stochastic online learning, and the deterministic online learning. 
Similarly, we also have three variations for REM, DLA, and PairD.

Most algorithms in the bandit learning family, however, are not as flexible as counterfactual learning algorithms.
Due to their needs to control and collect user clicks on multiple ranked lists for each query, algorithms like DBGD, MGD, and NSGD can only be used in online learning paradigms\footnote{\revised{Though it is possible to run DBGD offline with special strategies such as probabilistic interleaving, the corresponding results are too terrible to be used in any reasonable ranking systems~\cite{hofmann2013reusing}, so we ignore them in this paper.}}.
PDGD is an exception.
As described by Oosterhuis and De Rijke~\cite{oosterhuis2018differentiable}, PDGD requires no online result interleaving, which means that it is possible to adapt PDGD to offline learning by simply using $\pi_q$ from the offline data as the ranked lists.
\revised{Specifically, we compute the weights of ranked lists in PDGD (i.e., $\rho(d_i, d_j, \pi_q)$ in Eq.~(\ref{equ:pdgd_weight})) using the probability of the logged ranked list under the Plackett-Luce model described in Eq.~(\ref{equ:stochastic_ranking}).}
Therefore, we have two variations of DBGD, MGD, and NSGD as DBGD$_{OnD}$, DBGD$_{OnS}$, MGD$_{OnD}$, MGD$_{OnS}$, NSGD$_{OnD}$, and NSGD$_{OnS}$, and three variations of PDGD as PDGD$_{Off}$, PDGD$_{OnD}$, and PDGD$_{OnS}$.





\subsection{Effect of Learning Paradigms}

\begin{figure}[t]
	\centering
	\includegraphics[width=4in]{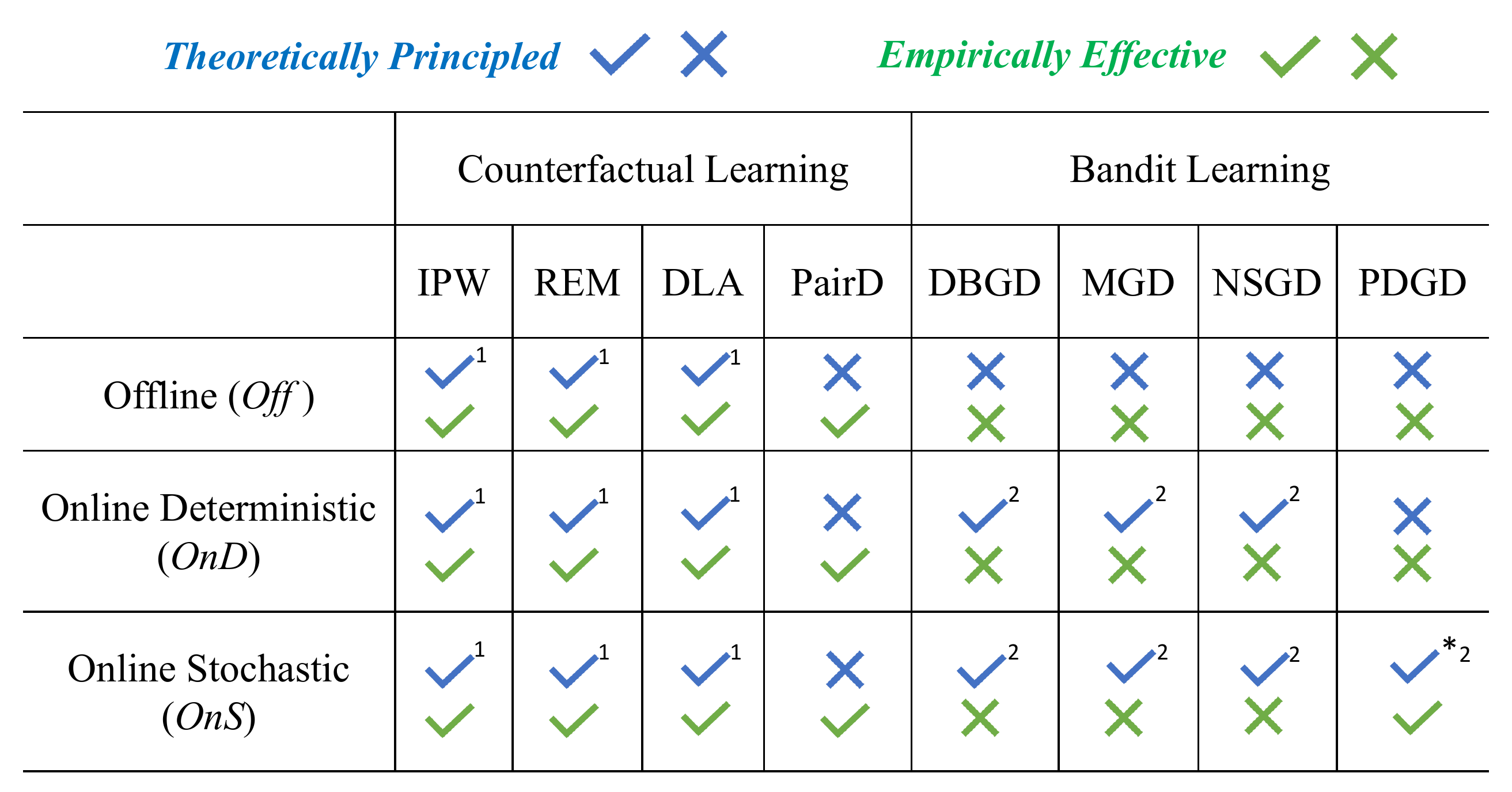}
	\caption{Unbiased learning-to-rank algorithms and learning paradigms. Theoretically principled algorithms under Theorem~\ref{def:unbiased_counterfactual} and Theorem~\ref{def:unbiased_bandit} are indicated with ``$^1$'' and ``$^2$'', and PDGD with Online Stochastic Learning is labeled with ``*'' because it is only partially unbiased given specific restrictions discussed in Section~\ref{sec:PDGD}.}
	\label{fig:paradigm_effect}
\end{figure}

The comparison of offline learning paradigms and online learning paradigms has received considerable attention in the studies of learning to rank~\cite{liu2009learning}.
Empirically, it is widely believed that, given same types of ranking functions, online learning paradigms are likely to produce more effective ranking models than offline learning as the former can alleviate the problem of selection bias~\cite{wang2016learning} and collect direct feedback on the current state of the ranking function~\cite{oosterhuis2018differentiable}.
In contrast, offline learning often has less parameter variance than online learning~\cite{wang2019variance} and much less cost on user experience and system development in practice.
In unbiased learning to rank, however, the effect of learning paradigms varies from algorithms to algorithms.
A illustration of how different learning paradigms (i.e., \textit{Off}, \textit{OnS}, and \textit{OnD}) affect the theoretical foundations and empirical effectiveness of each unbiased learning to rank algorithms is shown in Figure~\ref{fig:paradigm_effect}.  

In counterfactual learning, algorithms like IPW and REM have no assumption on the distribution of $\pi_q$, so the change of learning paradigms will have no effect on their effectiveness in theory.
For DLA, however, things are more complicated.
Because the proof in Eq.~(\ref{equ:irw_expectation}) relies on the assumption that $\mathbb{E}_{\mathbf{r}}[r_d]$ is equal to $P(r_d=1)$ for each result position, the unbiasness of DLA is not guaranteed when the backend model of $\pi_q$ keeps changing. 
However, when the training of $f_{\theta}$ is close to convergence or the learning rate is small enough, $\mathbb{E}_{\mathbf{r}}[r_d]$ would be stable and thus equal or similar to $P(r_d=1)$ in stochastic or deterministic online learning.
Thus, applying online learning to DLA may hurt its robustness in training but still achieve good results in the end.
We ignore the discussion of PairD here because it is not theoretically principled.


In bandit learning, most algorithms are not applicable to or not theoretically principled in offline learning paradigms because $\pi_q$ is fixed in offline data and most bandit learning algorithms require the control of $\pi_q$ in order to achieve unbiasness.
Also, even in online settings, the effectiveness of bandit learning algorithms could be significantly affected by the sampling strategies of $\pi_q$.
For instance, while DBGD-based methods are theoretically principled in both stochastic online learning and deterministic online learning environments, they generally explore less rankings in deterministic online learning and have larger variance in stochastic online learning. 
How these affect the final performance of the algorithms significantly depend on parameter initialization, hyper-parameter settings, etc.
While PDGD doesn't involve any result interleaving in the training process, it is partially unbiased only when $\pi_q$ is strictly sampled based on $f_{\theta}$ in stochastic manners.
Any disturbance on the distribution of $\pi_q$ (e.g., change from \textit{OnS} to \textit{OnD}) would hurt the theoretical foundation as well as the performance of PDGD.








\section{Empirical Experiments}\label{sec:exp}


\revised{In order to answer the second research question of this paper (i.e., RQ2)}, in this section, we present our empirical analysis on unbiased learning-to-rank algorithms.
Specifically, we conduct experiments with both synthetic and real click data on public available learning-to-rank benchmarks to evaluate the effect of learning paradigms and the performance of existing unbiased learning-to-rank algorithms.

\subsection{Experiments with Synthetic Data}\label{sec:simluation_exp}


\subsubsection{Datasets and Simulation Setup}\label{sec:dataset}

To fully test unbiased learning-to-rank algorithms with different learning paradigms, we conducted experiments using synthetic click data derived from Yahoo! Learning to Rank Collection (set 1)\footnote{\url{http://webscope.sandbox.yahoo.com}} and Istella-S\footnote{\url{http://quickrank.isti.cnr.it/istella-dataset/}}~\cite{lucchese2016post}.
Yahoo! LETOR dataset contains 29,921 queries and 701k documents (approximately 23 documents per query) sampled from a commercial English search engine . 
Each query-document pair is represented with 700 features and annotated with 5-level relevance judgments from 0 (i.e. \textit{irrelevant}) to 4 (i.e. \textit{perfectly relevant}).
Istella-S dataset is composed of 33,018 queries and 3,408k documents (approximately 100 documents per query) sampled from a commercial Italian search engine.
Each query-document pair in Istella-S is represetned with 220 features and annotated with 5-level relevance judgments.
Due to privacy concerns, no click data is released on these dataset.
Therefore, we simulate click data following the methodology used by previous studies~\cite{joachims2017unbiased,ai2018unbiased,hu2019unbiased}.
Specifically, we sampled the probability of examination on a document $d$ as
$$
P(o_d=1) = P(o_i=1) = \nu_{i}^{\eta}
$$
where $i$ is the position of $d$ in the displayed ranked list $\pi_q$, $\nu_i$ is the examination probability on $i$ estimated by eye-tracking studies~\cite{joachims2005accurately} \revised{(as shown in Table~\ref{tab:exam_prob})}, and $\eta$ is a hyper-parameter that controls the severity of position bias.
We sampled the probability of $d$ being perceived as relevant as
$$
P(r_d=1) = \epsilon + (1-\epsilon)\frac{2^y-1}{2^{4}-1}
$$
where $y$ is the annotated relevance label of $d$ and $\epsilon$ is a hyper-parameter controlling the probability of noisy clicks. 
Then, we simulated user behavior by generating synthetic clicks according to Eq.~(\ref{equ:click_assumption}).
For simplicity, we fixed $\eta=1.0$ and $\epsilon=0.1$ unless stated otherwise.
It is worth noting that all experiments in this paper are conducted in environments with selection bias~\cite{wang2016learning}, which means that users can only see and click on the top 10 retrieved results for each query. 
Depending on the number of candidate documents for each query, the effect of selection bias could vary on different queries and different datasets.
We briefly discuss it in Section~\ref{sec:simluation_exp}.  
We ignore the analysis of unbiased learning to rank without selection bias~\cite{Jagerman:2019:MIC:3331184.3331269} as it is unrealistic in practice.

\begin{table}[t]
	\caption{The value of examination probabilities $\nu_{i}$ on each result position $i$ reported by Joachims et al.~\cite{joachims2005accurately} through eye-tracking experiments.
	}
	\centering
	\small	
	\vspace{-5pt}
	\scalebox{1}{
		\begin{tabular}{ c || c | c | c | c | c | c | c  | c | c | c     } 
			\hline \hline
			Position $i$ & 1 & 2 & 3& 4 & 5 & 6 & 7 & 8 & 9 & 10 \\ \hline \hline
			Exam. Probability $\nu_{i}$ & 0.68 & 0.61 & 0.48 & 0.34 & 0.28 & 0.20 & 0.11 & 0.10 & 0.08 & 0.06 \\ 
			\hline \hline
		\end{tabular}
	}
	\label{tab:exam_prob}
\end{table}

\subsubsection{Model setup and evaluation}

Our goal is to conduct a fair comparison of the unbiasness of different unbiased learning-to-rank algorithms with different learning paradigms. 
Therefore, in each experiment, we use a single type of ranking models for all unbiased learning-to-rank algorithms. 
Specifically, we tested two types of ranking models separately, which are the multiple-layer perceptron network (MLP) with non-linear ELU activation functions and the linear regression model. 
The MLP we used has three hidden layers (with 512, 256, and 128 neurons) and batch normalization on each layer before activation.
The ELU activation function is defined as 
$$
ELU(x) = \begin{cases} x, & \text{if }x \geq 0 \\ e^{x}-1, & \text{if } x<0 \end{cases}
$$
The linear regression model we used could be treated as a MLP model with no hidden layer and no activation function.
We implement the local loss function $l$ as pairwise cross entropy loss~\cite{burges2005learning} except for REM (which uses a sigmoid loss in the EM algorithm)~\cite{wang2018position} and DLA (which uses a softmax loss for dual learning)~\cite{ai2018unbiased}. 
We used online EM~\cite{cappe2009line} for EM algorithms and tuned learning rates from 0.01 to 0.05 for each unbiased learning-to-rank algorithm.
We set batch size as 256 and trained each algorithm for 10k steps.
For offline learning, we created a synthetic production model by training a Ranking SVM model~\cite{joachims2006training} with 1\% data randomly sampled from the original training set\footnote{\revised{The production ranker is built with the open-source tool from \url{http://www.cs.cornell.edu/people/tj/svm_light/svm_rank.html} using the default settings.}}. 
The production model is used to generate the ranked lists in offline click logs.
\revised{Also, for IPW, we conducted a separate online result randomization experiments with click simulation ($\eta=1.0$ and $\epsilon=0.1$) to estimate the inverse propensity weights. 
We repeatedly randomized the results shown in a million of sessions and collect simulated user clicks to the examination probability of each result position with Eq.~\ref{equ:random}.
Note that such randomization experiments could significantly increase the cost of IPW in practice.
}
For the reproducibility of our experiments and the future studies on related topics, we created an Unbiased Learning To Rank Algorithm (\textit{ULTRA}) toolbox that includes all the algorithms and experiment settings reported in this paper.
We will release the link to ULTRA after the publishment of this paper.

For evaluation, we trained and tested all models with the predefined training, validation, and test data in Yahoo! and Istella-S.
We used two standard ranking metrics -- the normalized Discounted Cumulative Gain (nDCG)~\cite{jarvelin2002cumulated} and the Expected Reciprocal Rank (ERR)~\cite{chapelle2009expected} to evaluate the performance of each algorithm.
NDCG is constructed based on the theory of information gain while ERR is built based on the model of user satisfaction in Web search.
Ranking models are selected according to their nDCG@10 \revised{(computed with true relevance labels)} on the validation data in training.
Each experiment are repeated for 5 times, and we compute the average metric value on top 1, 3, 5 and 10 results.
Significant test is conducted based on the Fisher randomization test~\cite{smucker2007comparison} with $p\leq 0.05$.
The experiment results on Yahoo! and Istella-S with MLP and linear regression models are shown in Table~\ref{tab:Yahoo_dnn}, \ref{tab:Yahoo_linear}, \ref{tab:istella_dnn}, and \ref{tab:istella_linear}.






\begin{table}[t]
	\caption{Comparison of unbiased learning-to-rank (ULTR) algorithms with different learning paradigms on \textbf{Yahoo!} LETOR data using \textbf{multiple-layer perceptron networks} (MLP) as ranking models. Significant improvements or degradations with respect to NA are indicated with $+/-$ in the Fisher randomization test~\cite{smucker2007comparison} with $p \leq 0.05$. The best performance is highlighted in boldface.
	}
	\begin{subtable}[t]{1.0\textwidth}
	\centering
		\small	
		\vspace{-5pt}
		\caption{Performance of unbiased learning-to-rank algorithms with offline learning (Off) on Yahoo! LETOR data.}
		\scalebox{0.88}{
			\begin{tabular}{ p{0.15\textwidth}| c || c | c | c | c | c | c  | c | c     } 
				\hline
				\multicolumn{10}{c}{Offline Learning (Off)}  \\ \hline
				\hline
				\multicolumn{2}{c||}{} &  nDCG@1 & ERR@1 & nDCG@3 & ERR@3 & nDCG@5 & ERR@5 & nDCG@10 & ERR@10 \\ \hline
				\hline
				\multirow{4}{0.15\textwidth}{Counterfactual Learning Family} 
				& IPW & 0.682$^+$ & 0.347$^+$ & 0.685$^+$ & 0.425$^+$ & \textbf{0.708}$^+$ & 0.447$^+$ & \textbf{0.755}$^+$ & 0.462$^+$ \\ \cline{2-10}
				& REM & 0.673$^+$ & 0.347$^+$ & 0.677$^+$ & 0.424$^+$ & 0.698$^+$ & 0.445$^+$ & 0.745$^+$ & 0.460$^+$ \\ \cline{2-10}
				& DLA & \textbf{0.684}$^+$ & \textbf{0.350}$^+$ & \textbf{0.686}$^+$ & \textbf{0.427}$^+$ & 0.707$^+$ & \textbf{0.449}$^+$ & 0.754$^+$ & \textbf{0.464}$^+$ \\ \cline{2-10}
				& PairD & 0.658$^-$ & 0.335 & 0.662$^-$ & 0.413 & 0.686 & 0.436 & 0.737 & 0.451 \\ \hline
				\hline
				\multirow{4}{0.15\textwidth}{Bandit Learning Family} 
				& DBGD &  - & - & - & - & - & - & - & - \\ \cline{2-10}
				& MGD &  - & - & - & - & - & - & - & - \\ \cline{2-10}
				& NSGD &  - & - & - & - & - & - & - & - \\ \cline{2-10}
				& PDGD & 0.334$^-$ & 0.127$^-$ & 0.392$^-$ & 0.206$^-$ & 0.443$^-$ & 0.238$^-$ & 0.533$^-$ & 0.263$^-$ \\ \hline
				\hline
				\multirow{2}{0.15\textwidth}{Others} 
				& NA & 0.661 & 0.335 & 0.664 & 0.414 & 0.687 & 0.436 & 0.738 & 0.452 \\ \cline{2-10}
				& Prod. & 0.593$^-$ & 0.291$^-$ & 0.616$^-$ & 0.377$^-$ & 0.645$^-$ & 0.401$^-$ & 0.705$^-$ & 0.418$^-$ \\ \hline
				\hline
			\end{tabular}
		}
		
		\label{tab:Yahoo_dnn_off}
	\end{subtable}

	\begin{subtable}[t]{1.0\textwidth}
		\centering
		\small	
		\vspace{5pt}
		\caption{Performance of unbiased learning-to-rank algorithms with stochastic online learning (OnS) on Yahoo! LETOR data.}
		\scalebox{0.88}{
			\begin{tabular}{ p{0.15\textwidth}| c || c | c | c | c | c | c  | c | c     } 
				\hline
				\multicolumn{10}{c}{Stochastic Online Learning (OnS)}  \\ \hline
				\hline
				\multicolumn{2}{c||}{} &  nDCG@1 & ERR@1 & nDCG@3 & ERR@3 & nDCG@5 & ERR@5 & nDCG@10 & ERR@10 \\ \hline
				\hline
				\multirow{4}{0.15\textwidth}{Counterfactual Learning Family} 
				& IPW & 0.685$^+$ & 0.347$^+$ & 0.686$^+$ & 0.426$^+$ & 0.706$^+$ & 0.447$^+$ & 0.753$^+$ & 0.463$^+$ \\ \cline{2-10}
				& REM & 0.671 & 0.347$^+$ & 0.675$^+$ & 0.423$^+$ & 0.696$^+$ & 0.445$^+$ & 0.744$^+$ & 0.460$^+$ \\ \cline{2-10}
				& DLA & 0.676$^+$ & 0.346$^+$ & 0.680$^+$ & 0.423$^+$ & 0.702$^+$ & 0.445$^+$ & 0.750$^+$ & 0.461$^+$ \\ \cline{2-10}
				& PairD & 0.678$^+$ & 0.346$^+$ & 0.679$^+$ & 0.423$^+$ & 0.701$^+$ & 0.445$^+$ & 0.749$^+$ & 0.460$^+$ \\ \hline
				\hline
				\multirow{4}{0.15\textwidth}{Bandit Learning Family} 
				& DBGD & 0.411$^-$ & 0.163$^-$ & 0.471$^-$ & 0.257$^-$ & 0.520$^-$ & 0.289$^-$ & 0.604$^-$ & 0.313$^-$ \\ \cline{2-10}
				& MGD & 0.426$^-$ & 0.178$^-$ & 0.482$^-$ & 0.272$^-$ & 0.530$^-$ & 0.302$^-$ & 0.611$^-$ & 0.324$^-$ \\ \cline{2-10}
				& NSGD & 0.433$^-$ & 0.179$^-$ & 0.495$^-$ & 0.277$^-$ & 0.544$^-$ & 0.308$^-$ & 0.623$^-$ & 0.329$^-$ \\ \cline{2-10}
				& PDGD & \textbf{0.687}$^+$ & \textbf{0.350}$^+$ & \textbf{0.689}$^+$ & \textbf{0.428}$^+$ & \textbf{0.709}$^+$ & \textbf{0.449}$^+$ & \textbf{0.756}$^+$ & \textbf{0.464}$^+$ \\ \hline
				\hline
				Others 
				& NA & 0.670 & 0.345 & 0.670 & 0.421 & 0.690 & 0.443 & 0.740 & 0.458 \\ \hline
				\hline
			\end{tabular}
		}
		
		\label{tab:Yahoo_dnn_ons}
	\end{subtable}

	\begin{subtable}[t]{1.0\textwidth}
		\centering
		\small	
		\vspace{5pt}
		\caption{Performance of unbiased learning-to-rank algorithms with deterministic online learning (OnD) on Yahoo! LETOR data.}
		\scalebox{0.88}{
			\begin{tabular}{ p{0.15\textwidth}| c || c | c | c | c | c | c  | c | c     } 
				\hline
				\multicolumn{10}{c}{Deterministic Online Learning (OnD)}  \\ \hline
				\hline
				\multicolumn{2}{c||}{} &  nDCG@1 & ERR@1 & nDCG@3 & ERR@3 & nDCG@5 & ERR@5 & nDCG@10 & ERR@10 \\ \hline
				\hline
				\multirow{4}{0.15\textwidth}{Counterfactual Learning Family} 
				& IPW & \textbf{0.681}$^+$ & 0.347$^+$ & \textbf{0.681}$^+$ & \textbf{0.425}$^+$ & \textbf{0.701}$^+$ & 0.446$^+$ & \textbf{0.749}$^+$ & \textbf{0.462}$^+$ \\ \cline{2-10}
				& REM & 0.680$^+$ & \textbf{0.350}$^+$ & 0.680$^+$ & \textbf{0.425}$^+$ & 0.700$^+$ & \textbf{0.447}$^+$ & 0.748$^+$ & \textbf{0.462}$^+$ \\ \cline{2-10}
				& DLA & 0.675$^+$ & 0.347$^+$ & 0.679$^+$ & 0.424$^+$ & \textbf{0.701}$^+$ & 0.445$^+$ & \textbf{0.749}$^+$ & 0.461$^+$ \\ \cline{2-10}
				& PairD & 0.656$^-$ & 0.340 & 0.653$^-$ & 0.415$^-$ & 0.674$^-$ & 0.437$^-$ & 0.725$^-$ & 0.453$^-$ \\ \hline
				\hline
				\multirow{4}{0.15\textwidth}{Bandit Learning Family} 
				& DBGD & 0.406$^-$ & 0.165$^-$ & 0.469$^-$ & 0.260$^-$ & 0.519$^-$ & 0.292$^-$ & 0.602$^-$ & 0.314$^-$ \\ \cline{2-10}
				& MGD & 0.406$^-$ & 0.167$^-$ & 0.467$^-$ & 0.260$^-$ & 0.516$^-$ & 0.291$^-$ & 0.599$^-$ & 0.314$^-$ \\ \cline{2-10}
				& NSGD & 0.409$^-$ & 0.167$^-$ & 0.472$^-$ & 0.262$^-$ & 0.522$^-$ & 0.293$^-$ & 0.605$^-$ & 0.316$^-$ \\ \cline{2-10}
				& PDGD & 0.645$^-$ & 0.330$^-$ & 0.648$^-$ & 0.408$^-$ & 0.673$^-$ & 0.431$^-$ & 0.726$^-$ & 0.447$^-$ \\ \hline
				\hline
				Others 
				& NA & 0.665 & 0.341 & 0.662 & 0.417 & 0.684 & 0.439 & 0.734 & 0.455 \\ \hline
				\hline
			\end{tabular}
		}
		
		\label{tab:Yahoo_dnn_ond}
	\end{subtable}
	\label{tab:Yahoo_dnn}
\end{table}

\clearpage
\begin{table}[t]
	\caption{Comparison of unbiased learning-to-rank (ULTR) algorithms with different learning paradigms on \textbf{Yahoo!} LETOR data using \textbf{linear regression} functions as ranking models. Significant improvements or degradations with respect to NA are indicated with $+/-$ in the Fisher randomization test~\cite{smucker2007comparison} with $p \leq 0.05$. The best performance is highlighted in boldface.
	}
	\begin{subtable}[t]{1.0\textwidth}
		\centering
		\small	
		\vspace{-5pt}
		\caption{Performance of unbiased learning-to-rank algorithms with offline learning (Off) on Yahoo! LETOR data.}
		\scalebox{0.88}{
			\begin{tabular}{ p{0.15\textwidth}| c || c | c | c | c | c | c  | c | c     } 
				\hline
				\multicolumn{10}{c}{Offline Learning (Off)}  \\ \hline
				\hline
				\multicolumn{2}{c||}{} &  nDCG@1 & ERR@1 & nDCG@3 & ERR@3 & nDCG@5 & ERR@5 & nDCG@10 & ERR@10 \\ \hline
				\hline
				\multirow{4}{0.15\textwidth}{Counterfactual Learning Family} 
				& IPW & \textbf{0.675}$^+$ & 0.345$^+$ & \textbf{0.679}$^+$ & \textbf{0.423}$^+$ & \textbf{0.701}$^+$ & \textbf{0.445}$^+$ & \textbf{0.749}$^+$ & \textbf{0.460}$^+$ \\ \cline{2-10}
				& REM & 0.652 & 0.337$^+$ & 0.654$^-$ & 0.413$^+$ & 0.676$^-$ & 0.435 & 0.728$^-$ & 0.451$^+$ \\ \cline{2-10}
				& DLA & 0.672$^+$ & \textbf{0.346}$^+$ & 0.676$^+$ & \textbf{0.423}$^+$ & 0.697$^+$ & 0.444$^+$ & 0.746$^+$ & \textbf{0.460}$^+$ \\ \cline{2-10}
				& PairD & 0.653 & 0.332 & 0.659 & 0.412 & 0.683 & 0.434 & 0.735 & 0.450 \\ \hline
				\hline
				\multirow{4}{0.15\textwidth}{Bandit Learning Family} 
				& DBGD &  - & - & - & - & - & - & - & - \\ \cline{2-10}
				& MGD &  - & - & - & - & - & - & - & - \\ \cline{2-10}
				& NSGD &  - & - & - & - & - & - & - & - \\ \cline{2-10}
				& PDGD & 0.320$^-$ & 0.117$^-$ & 0.376$^-$ & 0.195$^-$ & 0.424$^-$ & 0.226$^-$ & 0.516$^-$ & 0.252$^-$ \\ \hline
				\hline
				\multirow{2}{0.15\textwidth}{Others} 
				& NA & 0.653 & 0.331 & 0.659 & 0.411 & 0.683 & 0.434 & 0.735 & 0.449 \\ \cline{2-10}
				& Prod. & 0.593$^-$ & 0.291$^-$ & 0.616$^-$ & 0.377$^-$ & 0.645$^-$ & 0.401$^-$ & 0.705$^-$ & 0.418$^-$ \\ \hline
				\hline
			\end{tabular}
		}
		
		\label{tab:Yahoo_linear_off}
	\end{subtable}
	
	\begin{subtable}[t]{1.0\textwidth}
		\centering
		\small	
		\vspace{5pt}
		\caption{Performance of unbiased learning-to-rank algorithms with stochastic online learning (OnS) on Yahoo! LETOR data.}
		\scalebox{0.88}{
			\begin{tabular}{ p{0.15\textwidth}| c || c | c | c | c | c | c  | c | c     } 
				\hline
				\multicolumn{10}{c}{Stochastic Online Learning (OnS)}  \\ \hline
				\hline
				\multicolumn{2}{c||}{} &  nDCG@1 & ERR@1 & nDCG@3 & ERR@3 & nDCG@5 & ERR@5 & nDCG@10 & ERR@10 \\ \hline
				\hline
				\multirow{4}{0.15\textwidth}{Counterfactual Learning Family} 
				& IPW & \textbf{0.683}$^+$ & \textbf{0.348} & 0.682$^+$ & \textbf{0.425} & \textbf{0.703}$^+$ & \textbf{0.447}$^+$ & \textbf{0.751}$^+$ & \textbf{0.462} \\ \cline{2-10}
				& REM & 0.661$^-$ & 0.343$^-$ & 0.663$^-$ & 0.418$^-$ & 0.685$^-$ & 0.440$^-$ & 0.736$^-$ & 0.456$^-$ \\ \cline{2-10}
				& DLA & 0.677 & 0.347 & 0.679 & 0.424 & 0.700 & 0.446 & 0.748 & 0.461 \\ \cline{2-10}
				& PairD & 0.666$^-$ & 0.341$^-$ & 0.669$^-$ & 0.419$^-$ & 0.691$^-$ & 0.441$^-$ & 0.741$^-$ & 0.457$^-$ \\ \hline
				\hline
				\multirow{4}{0.15\textwidth}{Bandit Learning Family} 
				& DBGD & 0.412$^-$ & 0.167$^-$ & 0.474$^-$ & 0.262$^-$ & 0.524$^-$ & 0.294$^-$ & 0.607$^-$ & 0.316$^-$ \\ \cline{2-10}
				& MGD & 0.396$^-$ & 0.157$^-$ & 0.455$^-$ & 0.248$^-$ & 0.505$^-$ & 0.280$^-$ & 0.591$^-$ & 0.304$^-$ \\ \cline{2-10}
				& NSGD & 0.434$^-$ & 0.165$^-$ & 0.488$^-$ & 0.260$^-$ & 0.536$^-$ & 0.292$^-$ & 0.619$^-$ & 0.315$^-$ \\ \cline{2-10}
				& PDGD & 0.682$^+$ & \textbf{0.348} & \textbf{0.683}$^+$ & \textbf{0.425} & \textbf{0.703}$^+$ & \textbf{0.447}$^+$ & 0.750$^+$ & \textbf{0.462} \\ \hline
				\hline
				Others 
				& NA & 0.678 & 0.347 & 0.679 & \textbf{0.425} & 0.700 & 0.446 & 0.748 & \textbf{0.462} \\ \hline
				\hline
			\end{tabular}
		}
		
		\label{tab:Yahoo_linear_ons}
	\end{subtable}
	
	\begin{subtable}[t]{1.0\textwidth}
		\centering
		\small	
		\vspace{5pt}
		\caption{Performance of unbiased learning-to-rank algorithms with deterministic online learning (OnD) on Yahoo! LETOR data.}
		\scalebox{0.88}{
			\begin{tabular}{ p{0.15\textwidth}| c || c | c | c | c | c | c  | c | c     } 
				\hline
				\multicolumn{10}{c}{Deterministic Online Learning (OnD)}  \\ \hline
				\hline
				\multicolumn{2}{c||}{} &  nDCG@1 & ERR@1 & nDCG@3 & ERR@3 & nDCG@5 & ERR@5 & nDCG@10 & ERR@10 \\ \hline
				\hline
				\multirow{4}{0.15\textwidth}{Counterfactual Learning Family} 
				& IPW & 0.677$^+$ & \textbf{0.347}$^+$ & 0.680$^+$ & \textbf{0.424}$^+$ & 0.700$^+$ & \textbf{0.446}$^+$ & 0.748$^+$ & \textbf{0.461}$^+$ \\ \cline{2-10}
				& REM & 0.668$^+$ & 0.345$^+$ & 0.668 & 0.421$^+$ & 0.689 & 0.442$^+$ & 0.738 & 0.458$^+$ \\ \cline{2-10}
				& DLA & \textbf{0.680}$^+$ & \textbf{0.347}$^+$ & \textbf{0.681}$^+$ & \textbf{0.424}$^+$ & \textbf{0.702}$^+$ & \textbf{0.446}$^+$ & \textbf{0.750}$^+$ & \textbf{0.461}$^+$ \\ \cline{2-10}
				& PairD & 0.657$^-$ & 0.340$^-$ & 0.656$^-$ & 0.416$^-$ & 0.680$^-$ & 0.438$^-$ & 0.731$^-$ & 0.454$^-$ \\ \hline
				\hline
				\multirow{4}{0.15\textwidth}{Bandit Learning Family} 
				& DBGD & 0.364$^-$ & 0.138$^-$ & 0.425$^-$ & 0.225$^-$ & 0.476$^-$ & 0.257$^-$ & 0.566$^-$ & 0.283$^-$ \\ \cline{2-10}
				& MGD & 0.418$^-$ & 0.172$^-$ & 0.480$^-$ & 0.268$^-$ & 0.530$^-$ & 0.299$^-$ & 0.611$^-$ & 0.321$^-$ \\ \cline{2-10}
				& NSGD & 0.424$^-$ & 0.163$^-$ & 0.482$^-$ & 0.258$^-$ & 0.530$^-$ & 0.290$^-$ & 0.614$^-$ & 0.314$^-$ \\ \cline{2-10}
				& PDGD & 0.657$^-$ & 0.340$^-$ & 0.656$^-$ & 0.416$^-$ & 0.680$^-$ & 0.438$^-$ & 0.731$^-$ & 0.454$^-$ \\ \hline
				\hline
				Others 
				& NA & 0.666 & 0.342 & 0.667 & 0.419 & 0.689 & 0.441 & 0.739 & 0.456 \\ \hline
				\hline
			\end{tabular}
		}
		
		\label{tab:Yahoo_linear_ond}
	\end{subtable}
	\label{tab:Yahoo_linear}
\end{table}

\clearpage
\begin{table*}[t]
	\caption{Comparison of unbiased learning-to-rank (ULTR) algorithms with different learning paradigms on \textbf{Istella-S} data using \textbf{multiple-layer perceptron networks} (MLP) as ranking models. Significant improvements or degradations with respect to the offline version of NA are indicated with $+/-$. Significant improvements or degradations with respect to NA are indicated with $+/-$ in the Fisher randomization test~\cite{smucker2007comparison} with $p \leq 0.05$. The best performance is highlighted in boldface.
	}
	\begin{subtable}[t]{1.0\textwidth}
		\centering
		\small	
		\vspace{-5pt}
		\caption{Performance of unbiased learning-to-rank algorithms with offline learning (Off) on Istella-S data.}
		\scalebox{0.88}{
			\begin{tabular}{ p{0.15\textwidth}| c || c | c | c | c | c | c  | c | c     } 
				\hline
				\multicolumn{10}{c}{Offline Learning (Off)}  \\ \hline
				\hline
				\multicolumn{2}{c||}{} &  nDCG@1 & ERR@1 & nDCG@3 & ERR@3 & nDCG@5 & ERR@5 & nDCG@10 & ERR@10 \\ \hline
				\hline
				\multirow{4}{0.15\textwidth}{Counterfactual Learning Family} 
				& IPW  & 0.662$^+$ & 0.591$^+$& 0.631$^+$ & 0.701$^+$ & \textbf{0.655}$^+$ & 0.717$^+$ & {0.714}$^+$& {0.724}$^+$  \\ \cline{2-10}
				& REM & 0.610$^-$ & {0.547}$^-$& {0.574}$^-$ & {0.657}$^-$ & {0.593}$^-$ & {0.675}$^-$ & {0.644}$^-$& {0.684}$^-$  \\ \cline{2-10}
				& DLA & \textbf{0.664}$^+$& \textbf{0.593}$^+$ & \textbf{0.632}$^+$  & \textbf{0.702}$^+$ & \textbf{0.655}$^+$ & \textbf{0.719}$^+$  & \textbf{0.715}$^+$& \textbf{0.725}$^+$ \\ \cline{2-10}
				& PairD & {0.623$^-$ }& 0.554$^-$ & {0.598}$^-$  & 0.672$^-$ & 0.625$^-$ & 0.691$^-$  & {0.691}$^-$& 0.699$^-$ \\ \hline
				\hline
				\multirow{4}{0.15\textwidth}{Bandit Learning Family} 
				& DBGD &  - & - & - & - & - & - & - & - \\ \cline{2-10}
				
				& MGD &  - & - & - & - & - & - & - & - \\ \cline{2-10}
				& NSGD &  - & - & - & - & - & - & - & - \\ \cline{2-10}
				
				& PDGD & 0.017$^-$ & 0.014$^-$ & 0.018$^-$ & 0.024$^-$ & 0.021$^-$ & 0.028$^-$ & 0.028$^-$ & 0.033$^-$ \\ \hline
				\hline
				\multirow{2}{0.15\textwidth}{Others} 
				& NA  & 0.633& 0.564  & 0.607 & 0.680 & 0.633& 0.698 & 0.697 & 0.705\\ \cline{2-10}
				& Prod. & 0.576$^-$ & 0.513$^-$  & 0.562$^-$ & 0.640$^-$ & 0.594$^-$ & 0.660$^-$ & 0.663$^-$& 0.669$^-$ \\ \hline
				\hline
			\end{tabular}
		}
		
		\label{tab:istella_dnn_off}
	\end{subtable}
	
	\begin{subtable}[t]{1.0\textwidth}
		\centering
		\small	
		\vspace{5pt}
		\caption{Performance of unbiased learning-to-rank algorithms with stochastic online learning (OnS) on Istella-S data.}
		\scalebox{0.88}{
			\begin{tabular}{ p{0.15\textwidth}| c || c | c | c | c | c | c  | c | c     } 
				\hline
				\multicolumn{10}{c}{Stochastic Online Learning (OnS)}  \\ \hline
				\hline
				\multicolumn{2}{c||}{} &  nDCG@1 & ERR@1 & nDCG@3 & ERR@3 & nDCG@5 & ERR@5 & nDCG@10 & ERR@10 \\ \hline
				\hline
				\multirow{4}{0.15\textwidth}{Counterfactual Learning Family} 
				& IPW & 0.657$^+$ & 0.587$^+$ & 0.627$^+$ & 0.697$^+$ & 0.650$^+$ & 0.714$^+$ & 0.710$^+$ & 0.721$^+$ \\ \cline{2-10}
				& REM & 0.644$^-$ & {0.574}$^-$& {0.612}$^-$ & {0.686}$^-$ & {0.636}$^-$ & {0.703}$^-$ & {0.699}$^-$& {0.711}$^-$  \\\cline{2-10}
				& DLA & 0.662$^+$ & 0.591$^+$ & 0.630$^+$ & 0.701$^+$ & 0.654$^+$ & 0.718$^+$ & 0.716$^+$ & 0.724$^+$ \\ \cline{2-10}
				& PairD & 0.660$^+$ & 0.589$^+$ & 0.628$^+$ & 0.698$^+$ & 0.652$^+$ & 0.715$^+$ & 0.715$^+$ & 0.722$^+$ \\ \hline
				\hline
				\multirow{4}{0.15\textwidth}{Bandit Learning Family} 
				& DBGD & 0.287$^-$ & 0.253$^-$ & 0.288$^-$ & 0.361$^-$ & 0.313$^-$ & 0.392$^-$ & 0.369$^-$ & 0.413$^-$\\ \cline{2-10}
				
				& MGD & 0.230$^-$ & 0.203$^-$ & 0.231$^-$ & 0.296$^-$ & 0.252$^-$ & 0.324$^-$ & 0.300$^-$ & 0.346$^-$ \\ \cline{2-10}
				& NSGD & 0.267$^-$ & 0.236$^-$ & 0.268$^-$ & 0.340$^-$ & 0.288$^-$ & 0.369$^-$ & 0.339$^-$ & 0.390$^-$ \\ \cline{2-10}
				
				& PDGD &\textbf{0.670}$^+$ & \textbf{0.598}$^+$ & \textbf{0.636}$^+$ & \textbf{0.706}$^+$ & \textbf{0.658}$^+$ & \textbf{0.722}$^+$ & \textbf{0.717}$^+$ & \textbf{0.729}$^+$ \\ \hline
				\hline
				Others 
				& NA & 0.652 & 0.582 & 0.623 & 0.693 &0.647 & 0.710 & 0.711 & 0.717 \\ \hline
				\hline
			\end{tabular}
		}
		
		\label{tab:istella_dnn_ons}
	\end{subtable}
	
	\begin{subtable}[t]{1.0\textwidth}
		\centering
		\small	
		\vspace{5pt}
		\caption{Performance of unbiased learning-to-rank algorithms with deterministic online learning (OnD) on Istella-S data.}
		\scalebox{0.88}{
			\begin{tabular}{ p{0.15\textwidth}| c || c | c | c | c | c | c  | c | c     } 
				\hline
				\multicolumn{10}{c}{Deterministic Online Learning (OnD)}  \\ \hline
				\hline
				\multicolumn{2}{c||}{} &  nDCG@1 & ERR@1 & nDCG@3 & ERR@3 & nDCG@5 & ERR@5 & nDCG@10 & ERR@10 \\ \hline
				\hline
				\multirow{4}{0.15\textwidth}{Counterfactual Learning Family} 
				& IPW & 0.663$^+$ & 0.592$^+$ & 0.633$^+$ & 0.702$^+$ & 0.657$^+$ & 0.719$^+$  & 0.717$^+$ & 0.726$^+$ \\ \cline{2-10}
				& REM  & 0.663$^+$ & 0.593$^+$ & 0.628$^+$ & 0.699$^+$ & 0.650$^+$ & 0.716$^+$ & 0.708$^+$ & 0.723$^+$  \\ \cline{2-10}
				& DLA & 0.668$^+$ & 0.596$^+$ & 0.633$^+$ & 0.705$^+$ & 0.656$^+$ & 0.721$^+$ & 0.715$^+$ & 0.728$^+$ \\ \cline{2-10}
				& PairD & \textbf{0.669}$^+$ & \textbf{0.598}$^+$ & \textbf{0.638}$^+$ & \textbf{0.707}$^+$ & \textbf{0.660}$^+$ & \textbf{0.723}$^+$ & \textbf{0.720}$^+$ & \textbf{0.729}$^+$ \\ \hline
				\hline
				\multirow{4}{0.15\textwidth}{Bandit Learning Family} 
				& DBGD & 0.310$^-$ & 0.276$^-$ & 0.305$^-$ & 0.384$^-$ & 0.327$^-$ & 0.412$^-$ & 0.378$^-$ & 0.431$^-$ \\ \cline{2-10}
				
				& MGD & 0.263$^-$ & 0.232$^-$ & 0.270$^-$ & 0.339$^-$ & 0.294$^-$ & 0.370$^-$ & 0.344$^-$ & 0.390$^-$\\ \cline{2-10}
				& NSGD & 0.253$^-$ & 0.222$^-$ & 0.252$^-$ & 0.318$^-$ & 0.271$^-$ & 0.344$^-$ & 0.314$^-$ & 0.364$^-$\\ \cline{2-10}
				
				& PDGD & 0.580$^-$ & 0.518$^-$ & 0.554$^-$ & 0.637$^-$ & 0.579$^-$ & 0.656$^-$ & 0.634$^-$ & 0.665$^-$ \\ \hline
				\hline
				Others 
				& NA & 0.654 & 0.584 & 0.623 & 0.695 & 0.645 & 0.711 & 0.704 & 0.718\\ \hline
				\hline
			\end{tabular}
		}
		
		\label{tab:istella_dnn_ond}
	\end{subtable}
	\label{tab:istella_dnn}
\end{table*}

\clearpage
\begin{table*}[t]
	\caption{Comparison of unbiased learning-to-rank (ULTR) algorithms with different learning paradigms on \textbf{Istella-S} data using \textbf{linear regression} functions as ranking models. Significant improvements or degradations with respect to the offline version of NA are indicated with $+/-$. Significant improvements or degradations with respect to NA are indicated with $+/-$ in the Fisher randomization test~\cite{smucker2007comparison} with $p \leq 0.05$. The best performance is highlighted in boldface.
	}
	\begin{subtable}[t]{1.0\textwidth}
		\centering
		\small	
		\vspace{-5pt}
		\caption{Performance of unbiased learning-to-rank algorithms with offline learning (Off) on Istella-S data.}
		\scalebox{0.88}{
			\begin{tabular}{ p{0.15\textwidth}| c || c | c | c | c | c | c  | c | c     } 
				\hline
				\multicolumn{10}{c}{Offline Learning (Off)}  \\ \hline
				\hline
				\multicolumn{2}{c||}{} &  nDCG@1 & ERR@1 & nDCG@3 & ERR@3 & nDCG@5 & ERR@5 & nDCG@10 & ERR@10 \\ \hline
				\hline
				\multirow{4}{0.15\textwidth}{Counterfactual Learning Family} 
				& IPW  & 0.628$^+$ & 0.563$^+$ & \textbf{0.601}$^+$ & 0.676$^+$ & 0.623$^+$ & 0.694$^+$ & 0.681$^+$ & 0.701$^+$  \\ \cline{2-10}
				& REM & 0.576$^+$ & 0.518$^+$ & 0.543$^+$ & 0.628$^+$ & 0.562$^+$ & 0.648$^+$ & 0.614$^+$ & 0.658$^+$   \\ \cline{2-10}
				& DLA & \textbf{0.632}$^+$ & \textbf{0.566}$^+$ & \textbf{0.601}$^+$ & \textbf{0.677}$^+$ & \textbf{0.624}$^+$ & \textbf{0.695}$^+$ & \textbf{0.682}$^+$ & \textbf{0.702}$^+$  \\ \cline{2-10}
				& PairD & 0.620$^+$ & 0.554$^+$ & 0.592$^+$ & 0.670$^+$ & 0.617$^+$ & 0.688$^+$ & 0.679$^+$ & 0.696$^+$ \\ \hline
				\hline
				\multirow{4}{0.15\textwidth}{Bandit Learning Family} 
				& DBGD &  - & - & - & - & - & - & - & - \\ \cline{2-10}
				
				& MGD &  - & - & - & - & - & - & - & - \\ \cline{2-10}
				& NSGD &  - & - & - & - & - & - & - & - \\ \cline{2-10}
				& PDGD  & 0.016$^-$ & 0.014$^-$ 
				& 0.017$^-$ & 0.023$^-$
				&0.020$^-$ & 0.028$^-$ 
				& 0.029$^-$ & 0.035$^-$ \\ \hline
				\hline
				\multirow{2}{0.15\textwidth}{Others} 
				& NA  & 0.570 &0.507
				& 0.544 & 0.620
				&0.570 & 0.650
				& 0.627 & 0.659 \\ \cline{2-10}
				& Prod. & 0.576$^+$ & 0.513$^+$  & 0.562$^+$ & 0.640$^+$ & 0.594$^+$ & 0.660$^+$ & 0.663$^+$& 0.669$^+$ \\ \hline
				\hline
			\end{tabular}
		}
		
		\label{tab:istella_linear_off}
	\end{subtable}
	
	\begin{subtable}[t]{1.0\textwidth}
		\centering
		\small	
		\vspace{5pt}
		\caption{Performance of unbiased learning-to-rank algorithms with stochastic online learning (OnS) on Istella-S data.}
		\scalebox{0.88}{
			\begin{tabular}{ p{0.15\textwidth}| c || c | c | c | c | c | c  | c | c     } 
				\hline
				\multicolumn{10}{c}{Stochastic Online Learning (OnS)}  \\ \hline
				\hline
				\multicolumn{2}{c||}{} &  nDCG@1 & ERR@1 & nDCG@3 & ERR@3 & nDCG@5 & ERR@5 & nDCG@10 & ERR@10 \\ \hline
				\hline
				\multirow{4}{0.15\textwidth}{Counterfactual Learning Family} 
				& IPW & 0.629$^+$ & 0.563$^+$ & 0.602$^+$ & 0.676$^+$ & \textbf{0.626}$^+$ & 0.694$^+$ & \textbf{0.683}$^+$ & 0.702$^+$ \\ \cline{2-10}
				& REM & 0.592$^+$ & 0.529$^+$ & 0.570$^+$ & 0.649$^+$ & 0.596$^+$ & 0.669$^+$ & 0.660$^+$ & 0.677$^+$ \\ \cline{2-10}
				& DLA & 0.621$^+$ & 0.556$^+$ & 0.595$^+$ & 0.671$^+$ & 0.619$^+$ & 0.689$^+$ & 0.681$^+$ & 0.697$^+$\\ \cline{2-10}
				& PairD & 0.601$^+$ & 0.537$^+$ & 0.579$^+$ & 0.657$^+$ & 0.605$^+$ & 0.676$^+$ & 0.668$^+$ & 0.684$^+$\\ \hline
				\hline
				\multirow{4}{0.15\textwidth}{Bandit Learning Family} 
				& DBGD & 0.280$^-$ & 0.248$^-$ & 0.275$^-$ & 0.350$^-$ & 0.295$^-$ & 0.377$^-$ & 0.342$^-$ & 0.396$^-$ \\ \cline{2-10}
				      
				& MGD & 0.277$^-$ & 0.247$^-$ & 0.264$^-$ & 0.343$^-$ & 0.281$^-$ & 0.368$^-$ & 0.321$^-$ & 0.386$^-$\\ \cline{2-10}		
				& NSGD & 0.224$^-$ & 0.198$^-$ & 0.223$^-$ & 0.287$^-$ & 0.241$^-$ & 0.314$^-$ & 0.280$^-$ & 0.335$^-$\\ \cline{2-10}	       				
				& PDGD & \textbf{0.633}$^+$ & \textbf{0.567}$^+$ & \textbf{0.603}$^+$ & \textbf{0.679}$^+$ & \textbf{0.626}$^+$ & \textbf{0.697}$^+$ & \textbf{0.683}$^+$ & \textbf{0.704}$^+$ \\ \hline
				\hline
				Others 
				& NA &0.571& 0.508 & 
				0.547 & 0.631 &
				0.573 & 0.651 & 
				0.634 & 0.660 \\ \hline
				\hline
			\end{tabular}
		}
		
		\label{tab:istella_linear_ons}
	\end{subtable}
	
	\begin{subtable}[t]{1.0\textwidth}
		\centering
		\small	
		\vspace{5pt}
		\caption{Performance of unbiased learning-to-rank algorithms with deterministic online learning (OnD) on Istella-S data.}
		\scalebox{0.88}{
			\begin{tabular}{ p{0.15\textwidth}| c || c | c | c | c | c | c  | c | c     } 
				\hline
				\multicolumn{10}{c}{Deterministic Online Learning (OnD)}  \\ \hline
				\hline
				\multicolumn{2}{c||}{} &  nDCG@1 & ERR@1 & nDCG@3 & ERR@3 & nDCG@5 & ERR@5 & nDCG@10 & ERR@10 \\ \hline
				\hline
				\multirow{4}{0.15\textwidth}{Counterfactual Learning Family} 
				& IPW & 0.635$^+$ & \textbf{0.569}$^+$ & 0.604$^+$ & \textbf{0.680}$^+$ & \textbf{0.628}$^+$ & \textbf{0.698}$^+$ & \textbf{0.686}$^+$ & \textbf{0.705}$^+$ \\ \cline{2-10}
				& REM & 0.629$^+$ & 0.563$^+$ & 0.596$^+$ & 0.674$^+$ & 0.620$^+$ & 0.692$^+$ & 0.677$^+$ & 0.700$^+$  \\ \cline{2-10}
				& DLA & \textbf{0.636}$^+$ & \textbf{0.569}$^+$ & \textbf{0.605}$^+$ & \textbf{0.680}$^+$ & \textbf{0.628}$^+$ & \textbf{0.698}$^+$ & 0.685$^+$ & \textbf{0.705}$^+$ \\ \cline{2-10}
				& PairD & 0.623$^+$ & 0.558$^+$ & 0.596$^+$ & 0.672$^+$ & 0.618$^+$ & 0.690$^+$ & 0.674$^+$ & 0.697$^+$ \\ \hline
				\hline
				\multirow{4}{0.15\textwidth}{Bandit Learning Family} 
				& DBGD & 0.324$^-$ & 0.288$^-$ & 0.315$^-$ & 0.397$^-$ & 0.335$^-$ & 0.424$^-$ & 0.382$^-$ & 0.443$^-$ \\ \cline{2-10}
				& MGD & 0.277$^-$ & 0.245$^-$ & 0.279$^-$ & 0.350$^-$ & 0.300$^-$ & 0.379$^-$ & 0.350$^-$ & 0.399$^-$\\ \cline{2-10}		
				& NSGD & 0.215$^-$ & 0.189$^-$ & 0.212$^-$ & 0.275$^-$ & 0.229$^-$ & 0.301$^-$ & 0.266$^-$ & 0.321$^-$\\ \cline{2-10}

				& PDGD & 0.595$^+$ & 0.532$^+$ & 0.572$^+$ & 0.651$^+$ & 0.598$^+$ & 0.670$^+$ & 0.658$^+$ & 0.679$^+$ \\ \hline
				\hline
				Others 
				& NA  & 0.452 & 0.406&
				0.405 & 0.506 & 
				0.411 & 0.528 & 
				0.448 & 0.543 \\ \hline
				\hline
			\end{tabular}
		}
		
		\label{tab:istella_linear_ond}
	\end{subtable}
	\label{tab:istella_linear}
\end{table*}

\clearpage

\subsection{How do learning paradigms affect algorithm effectiveness?}

Previous studies~\cite{liu2009learning,Jagerman:2019:MIC:3331184.3331269} argue that online learning can help ranking algorithms explore a greater parameter space than offline learning, especially when users can only see and click a limited number of results in each query (i.e., selection bias). 
Thus, it is believed that 

\vspace{5pt}
\textbf{H1}: \textit{Online learning paradigms are better than offline learning paradigms for unbiased learning to rank in environments with selection bias.} 
\vspace{5pt}




To test this hypothesis, we trained each unbiased learning-to-rank algorithm with both online and offline learning paradigms using MLP and linear regression functions as ranking models and summarize the results in Table~\ref{tab:Yahoo_dnn}, \ref{tab:Yahoo_linear}, \ref{tab:istella_dnn}, and \ref{tab:istella_linear}.
Algorithms in each table are grouped into three categories: (1) the counterfactual learning family, which includes IPW, REM, DLA, and PairD; (2) the bandit learning family, which includes DBGD, MGD, NSGD, and PDGD; (3) the production model (Prod.) used to generate offline click logs and the naive algorithm (NA) that directly trains ranking models with user clicks.
For simplicity, we only show the significant test results with respect to NA with offline learning (NA$_{off}$) in each table. 
However, it's worth noting that any differences larger or equal to 0.002 are statistically significant in the Fisher randomization test~\cite{smucker2007comparison} with $p \leq 0.05$. 

Table~\ref{tab:Yahoo_dnn} shows the retrieval performance of different learning paradigms and unbiased learning-to-rank algorithms on Yahoo! LETOR data using MLP as the ranking models.
As we can see in Table~\ref{tab:Yahoo_dnn}, IPW, DLA, and PDGD achieve the best performance among all the algorithms tested in our experiments.
DLA and IPW achieve the highest nDCG@10 in offline learning (Off) and deterministic online learning (OnD), while PDGD performed the best in stochastic online learning (OnS).
However, when comparing the best results of offline learning and online learning, we do not observe any significant differences.
The performance of the best offline model (i.e., DLA$_{Off}$) is mostly the same to the performance of the best online model (i.e., PDGD$_{OnS}$), which contradicts the hypothesis that online learning should be better than offline learning in environments with selection bias.

In Table~\ref{tab:Yahoo_linear}, we show the results of different algorithms on Yahoo! using linear regression functions as the ranking models.
Similar to Table~\ref{tab:Yahoo_dnn}, we observe that IPW, DLA, and PDGD usually performed the best in unbiased learning to rank given proper learning paradigms.
While linear regression models performed worse than MLP in most cases, the overall patterns and relative performance of different unbiased learning-to-rank algorithms with different learning paradigms are roughly the same.
For example, the best performance of models with online learning paradigms (i.e., OnS and OnD) is similar to the best performance of models with offline learning.
This, again, make the hypothesis of H1 seems questionable in our experiments. 



\begin{figure}
	\centering
	\begin{subfigure}{.5\textwidth}
		\centering
		\includegraphics[width=2.6in]{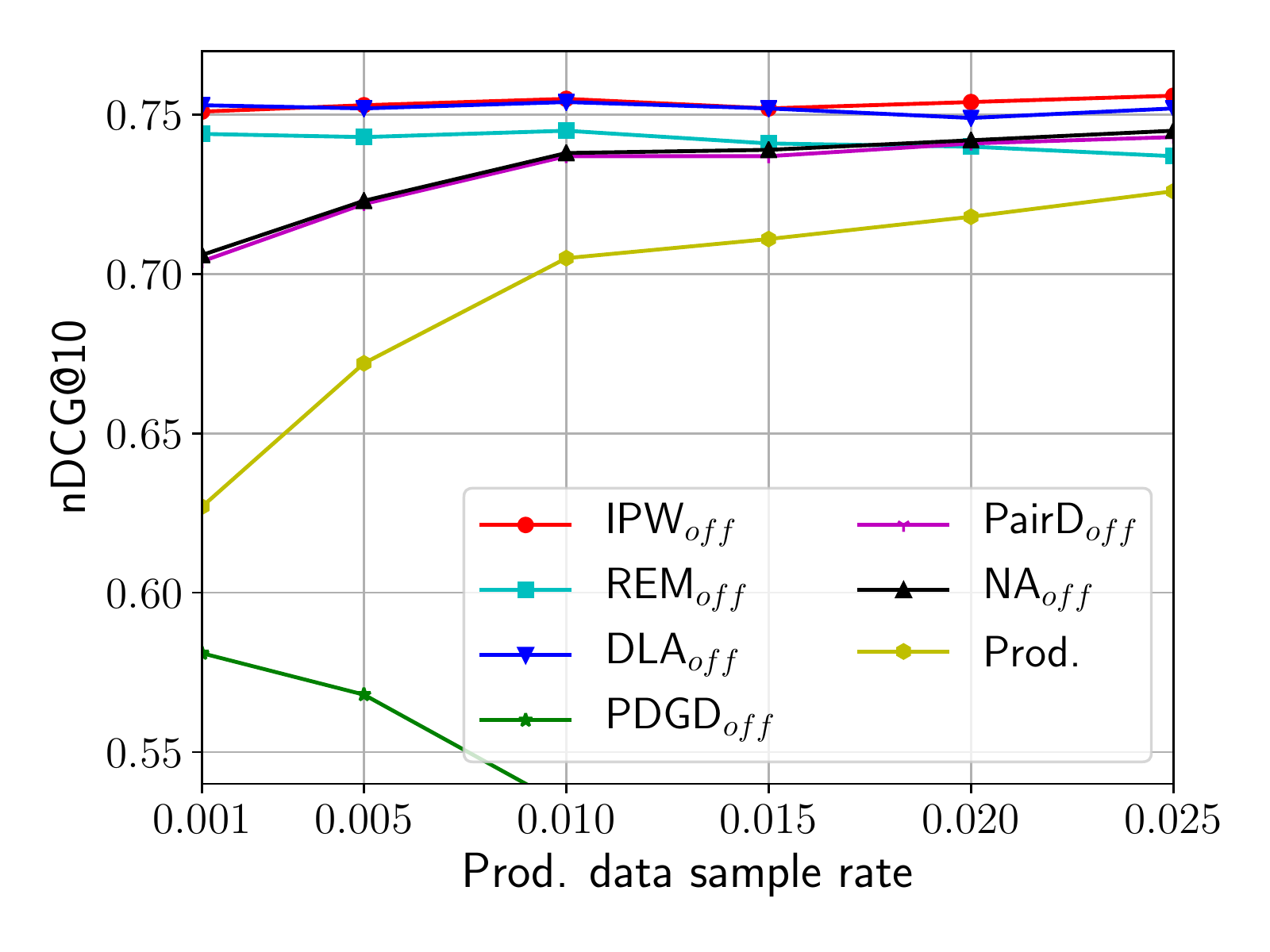} 
		\caption{Ranking with MLP}
		\label{fig:dnn_initial_ranker}
	\end{subfigure}%
	\begin{subfigure}{.5\textwidth}
		\centering
		\includegraphics[width=2.6in]{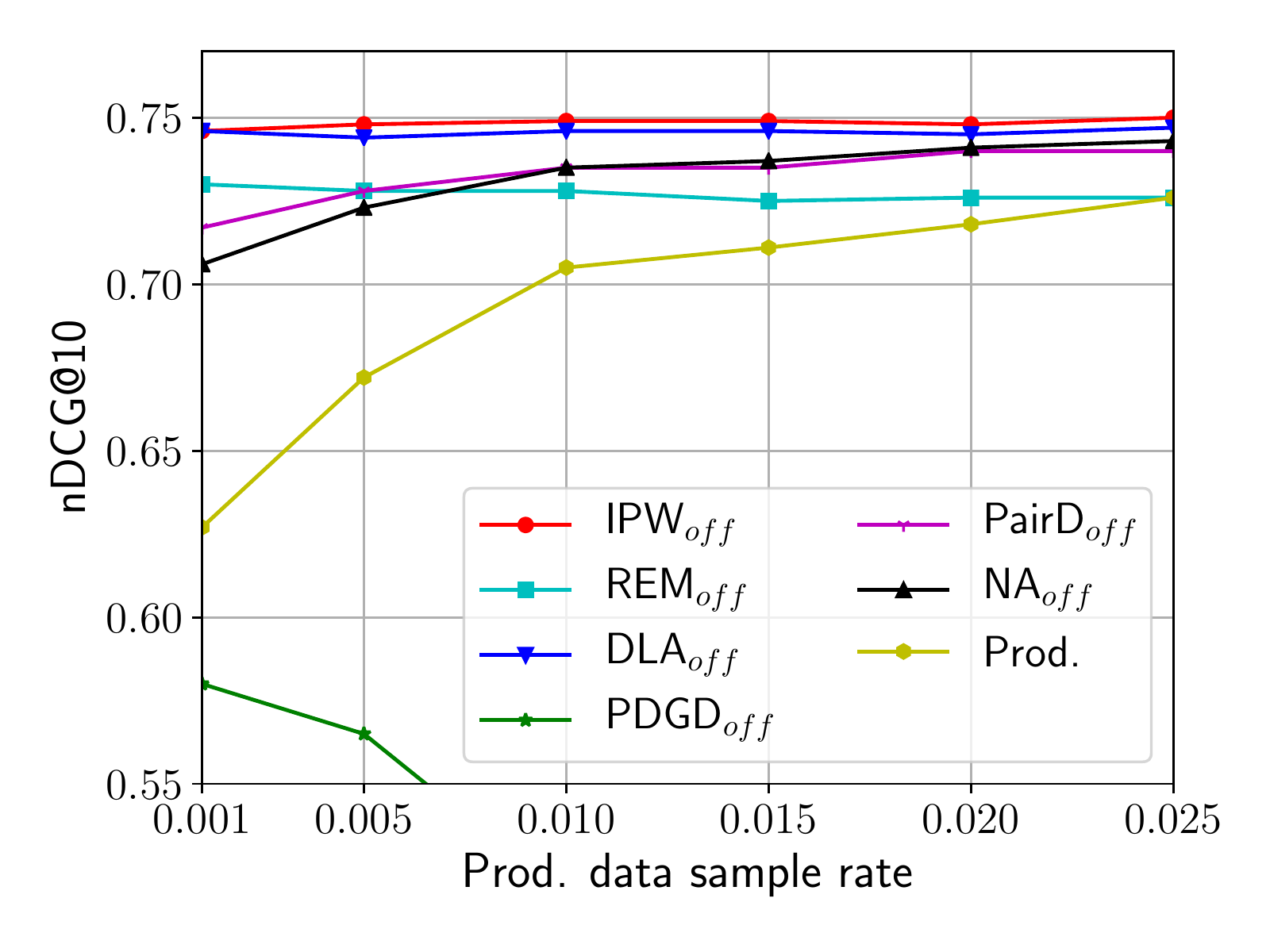}
		\caption{Ranking with linear regression}
		\label{fig:linear_initial_ranker}
	\end{subfigure}%
	\caption{Effect of production models for offline learning on Yahoo! LETOR data.}
	\label{fig:initial_ranker}%
\end{figure}

To validate this observation, we manipulate the production model in offline learning on Yahoo! LETOR data to check whether the performance of the logging systems would affect the performance of unbiased learning-to-rank algorithms.
Specifically, we change the size of the training data for the production model and plot the results in Figure~\ref{fig:initial_ranker}.
The x-axis of Figure~\ref{fig:initial_ranker} represents the proportion of sampled training data for the production model in the original training data of Yahoo! dataset, and the y-axis is the nDCG@10 of each algorithm in offline learning.
As depicted in Figure~\ref{fig:initial_ranker}, the performance of the production model (Prod.) increases when the size of the sampled training data increases.
Also, we observed that the performance of NA is positively correlated to the performance of the production system.
In contrast, the nDCGs of IPW, DLA and REM are relatively stable despite the change of production models.
This indicates that the high performance of IPW and DLA with offline learning in Table~\ref{tab:Yahoo_dnn}\&\ref{tab:Yahoo_linear} is not a coincidence.

Also, to further understand the effect of selection bias with respect to the effectiveness of offline and online learning paradigms, we jointly compare the experiments results on Yahoo! LETOR data and Istella-S. 
As mentioned in Section~\ref{sec:dataset}, the average number of candidate documents per query is approximately 23 and 100 in Yahoo! and Istella-S, respectively.
Since only the top 10 documents could be shown and clicked by users in our simulation experiments, it is reasonable to expect the problem of selection bias is more severe on Istella-S than Yahoo!.
In our experiments, we observe slightly different result patterns on Yahoo! and Istella-S data when comparing the performance of offline learning and online learning.
As shown in the result tables, the best performance of offline learning is usually achieved by counterfactual learning algorithms (e.g., IPW and DLA), and the best performance of online learning is usually achieved by bandit learning algorithms (i.e., PDGD with OnS).
When comparing the best offline models with the best online models on Yahoo!, we do not observe any evidence showing that online learning could achieve better performance than offline learning in unbiased learning to rank;
When comparing the best offline models with the best online models on Istella-S, we observe that unbiased learning-to-rank algorithms with online learning paradigms could achieve slight performance improvements (usually less than 0.5\%) than their offline versions.
This means that severe selection bias could have effect on the relative performance of offline learning and online learning, through the effect is minor in our experiments.

In fact, the phenomenon that offline unbiased learning to rank could achieve similar performance of online unbiased learning to rank is not surprising.
As shown in Eq.~(\ref{equ:cl_goal}), the unbiasness of counterfactual learning algorithms does not concern about the distribution of the displayed ranked list $\pi_q$. 
Theoretically speaking, methods such as IPW and DLA are guaranteed to find the best unbiased ranking models no matter how $\pi_q$ is created. 
As shown in Table~\ref{tab:Yahoo_dnn}\&\ref{tab:Yahoo_linear}, we indeed observe similar performance for the best offline model and the best online model.
\revised{Therefore, when the logging system of offline data is reasonable good (e.g., Prod. in Figure~\ref{fig:initial_ranker}) and the selection bias is not extraordinary severe, the effectiveness of unbiased learning to rank algorithms with offline learning paradigms is on par with those with online learning paradigms, which means that the hypothesis \textbf{H1} is not true from this perspective. 
However, please note that this does not indicate that online learning has no advantages over offline learning. 
There are many characteristics of online and offline optimization paradigms that haven't been touched in this study, such as convergence rate, robustness to local minimums, etc.   
These topics are beyond the scoop of this paper and we leave them for future studies.
}


\subsection{How do learning paradigms affect algorithm robustness?}
As discussed in Section~\ref{sec:theory} and \ref{sec:deployment}, the unbiasness of most counterfactual learning algorithms is independent from the distribution of displayed ranked list (i.e., $P( \pi_q|q)$), and the effectiveness of bandit learning algorithms is invariant to the inherited click bias in ranking loss $l(f_{\theta},\bm{c})$.
Therefore, we have the following hypotheses:

\vspace{5pt}
\textbf{H2}: \textit{The performance of counterfactual learning algorithms are invariant to learning paradigms (i.e., offline, stochastic online, or deterministic online), while the performance of bandit learning algorithms are sensitive to learning paradigms.} 

\vspace{2pt}
\textbf{H3}: \textit{With proper learning paradigms, bandit learning is more robust to variable click bias than counterfactual learning.} 
\vspace{5pt}

For the validation of \textbf{H2}, we report the performance of each unbiased learning-to-rank algorithm with different learning paradigms on Yahoo! in Table~\ref{tab:Yahoo_dnn}\&\ref{tab:Yahoo_linear}.
As shown in the tables, the performance of IPW and DLA is similar in offline learning, stochastic online learning, and deterministic online learning.
Their differences in different learning paradigms are less than 1\% in terms of nDCG and ERR. 
In contrast, we observe huge performance differences between the DBGD, MGD, NSGD, and PDGD using different learning paradigms. 
For example, the performance of PDGD using MLP in stochastic online learning is 42\% and 4\% better than PDGD in offline learning and deterministic online learning, respectively.
PDGD with offline or deterministic online learning actually performs worse than the naive algorithm that directly trains ranking models with clicks (i.e., NA) in Table~\ref{tab:Yahoo_dnn}\&\ref{tab:Yahoo_linear}.
As for DBGD and extend DBGD models (i.e., MGD and NSGD), we observe great variance and no consistent pattern with respect to their final performance with different learning paradigms.
Since they all have suboptimal performance comparing to other unbiased learning-to-rank algorithms, we ignore the further discussion of them for simplicity.


\begin{figure}
	\centering
	\begin{subfigure}{.5\textwidth}
		\centering
		\includegraphics[width=2.6in]{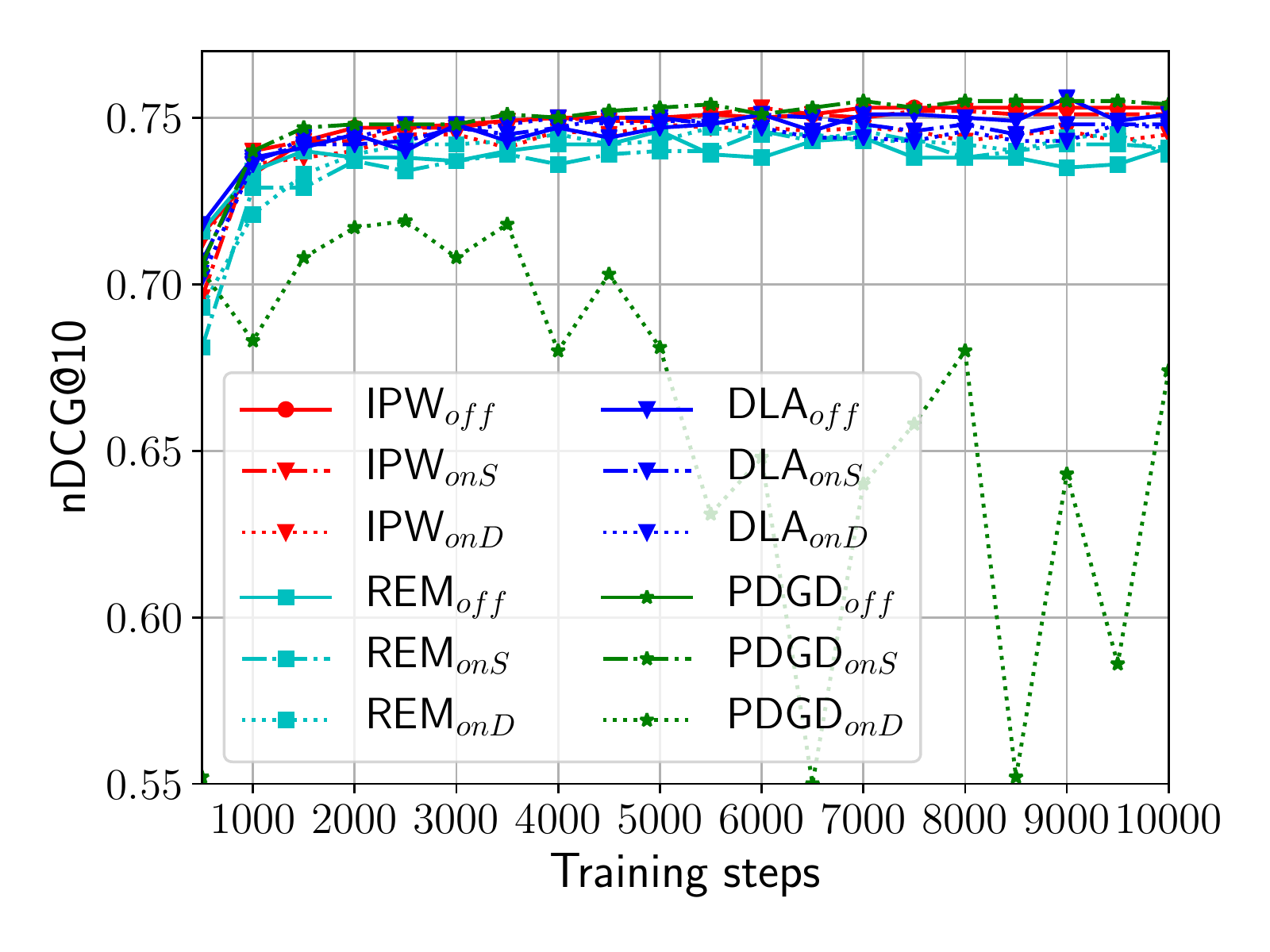} 
		\caption{Ranking with MLP}
		\label{fig:dnn_iteration}
	\end{subfigure}%
	\begin{subfigure}{.5\textwidth}
		\centering
		\includegraphics[width=2.6in]{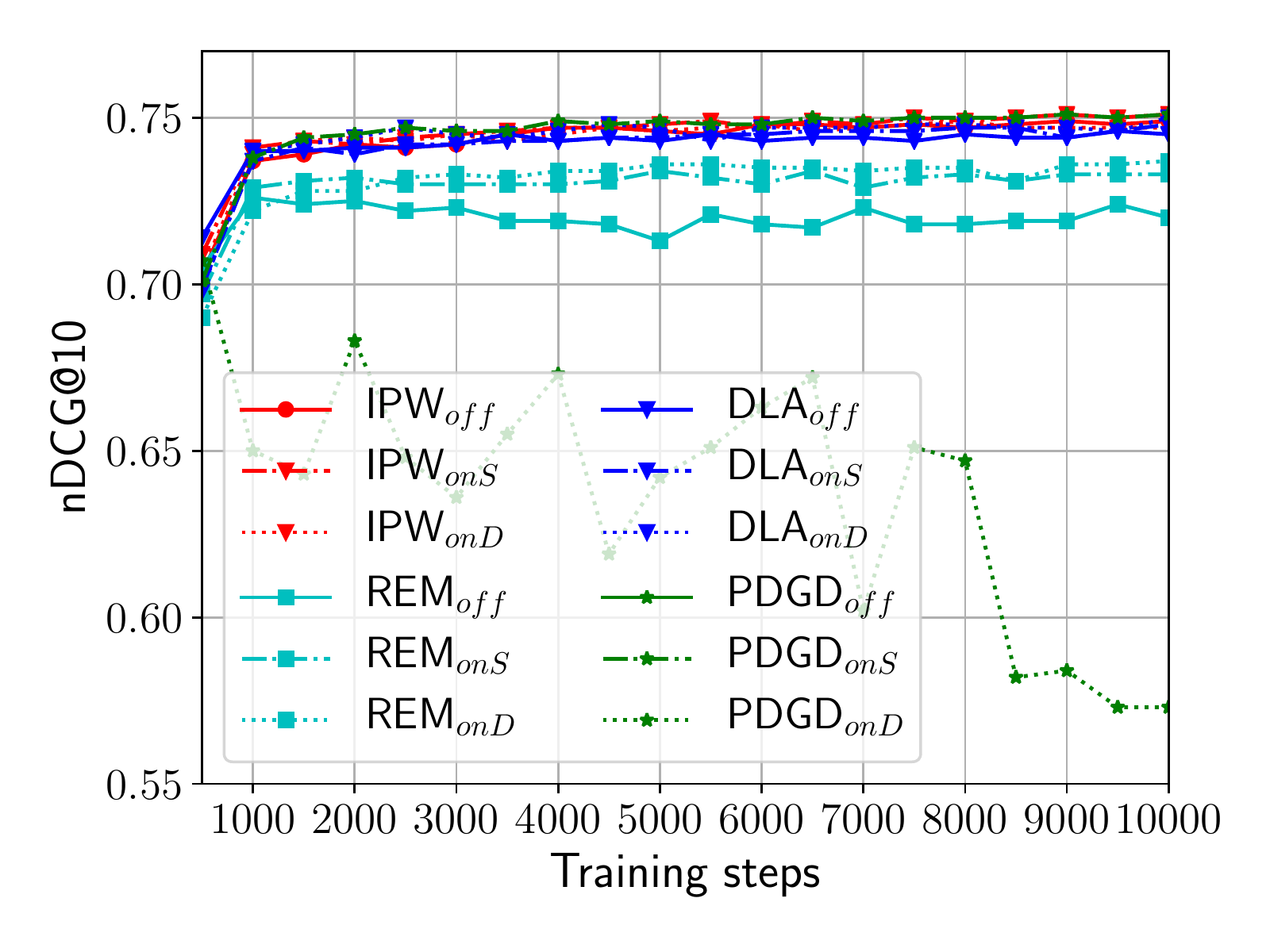}
		\caption{Ranking with linear regression}
		\label{fig:linear_iteration}
	\end{subfigure}%
	\caption{Test performance with respect to training steps on Yahoo! LETOR data.}
	\label{fig:iteration}%
\end{figure}

Figure~\ref{fig:iteration} plots the test performance of different algorithms in the training process.
As we can see in the figure, when the number of training steps increases, the learning curves of IPW, DLA, and REM with offline, stochastic online, and deterministic online learning are smooth and similar to each other.
The learning cure of PDGD with stochastic online learning (PDGD$_{onS}$) is also stable.
However, the performance of PDGD with deterministic online learning (PDGD$_{onD}$) shows a large variance in training and the performance of PDGD with offline learning (PDGD$_{off}$) is so bad that we can barely see it at the left bottom corner of Figure~\ref{fig:iteration}. 
Together with the results tables of Yahoo! and Istella-S, we conclude that bandit learning algorithms are indeed much more sensitive to learning paradigms than counterfactual learning algorithms, which means that \textbf{H2} is correct.




\begin{figure}
	\centering
	\begin{subfigure}{.5\textwidth}
		\centering
		\includegraphics[width=2.6in]{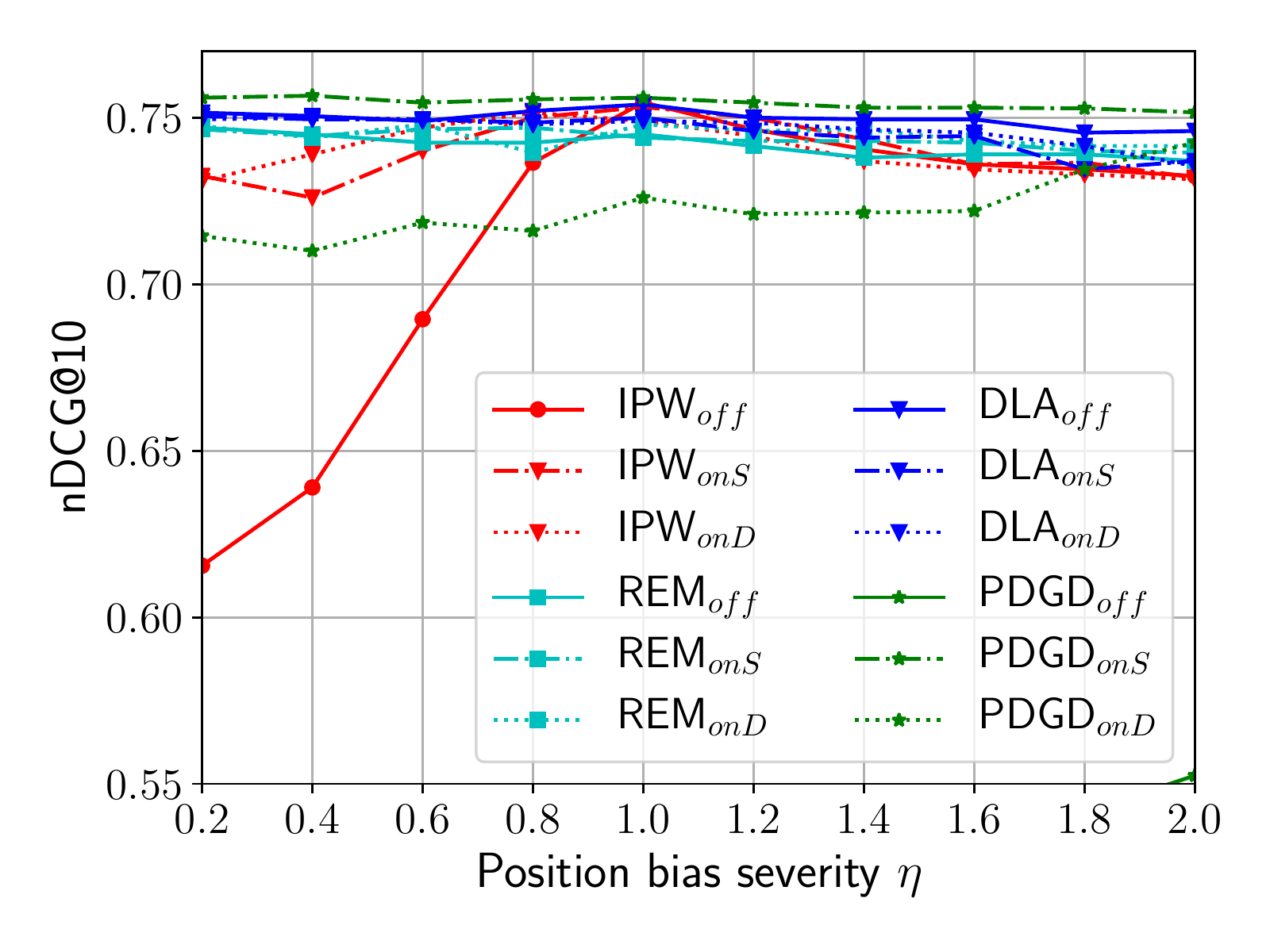} 
		\caption{Ranking with MLP}
		\label{fig:dnn_eta}
	\end{subfigure}%
	\begin{subfigure}{.5\textwidth}
		\centering
		\includegraphics[width=2.6in]{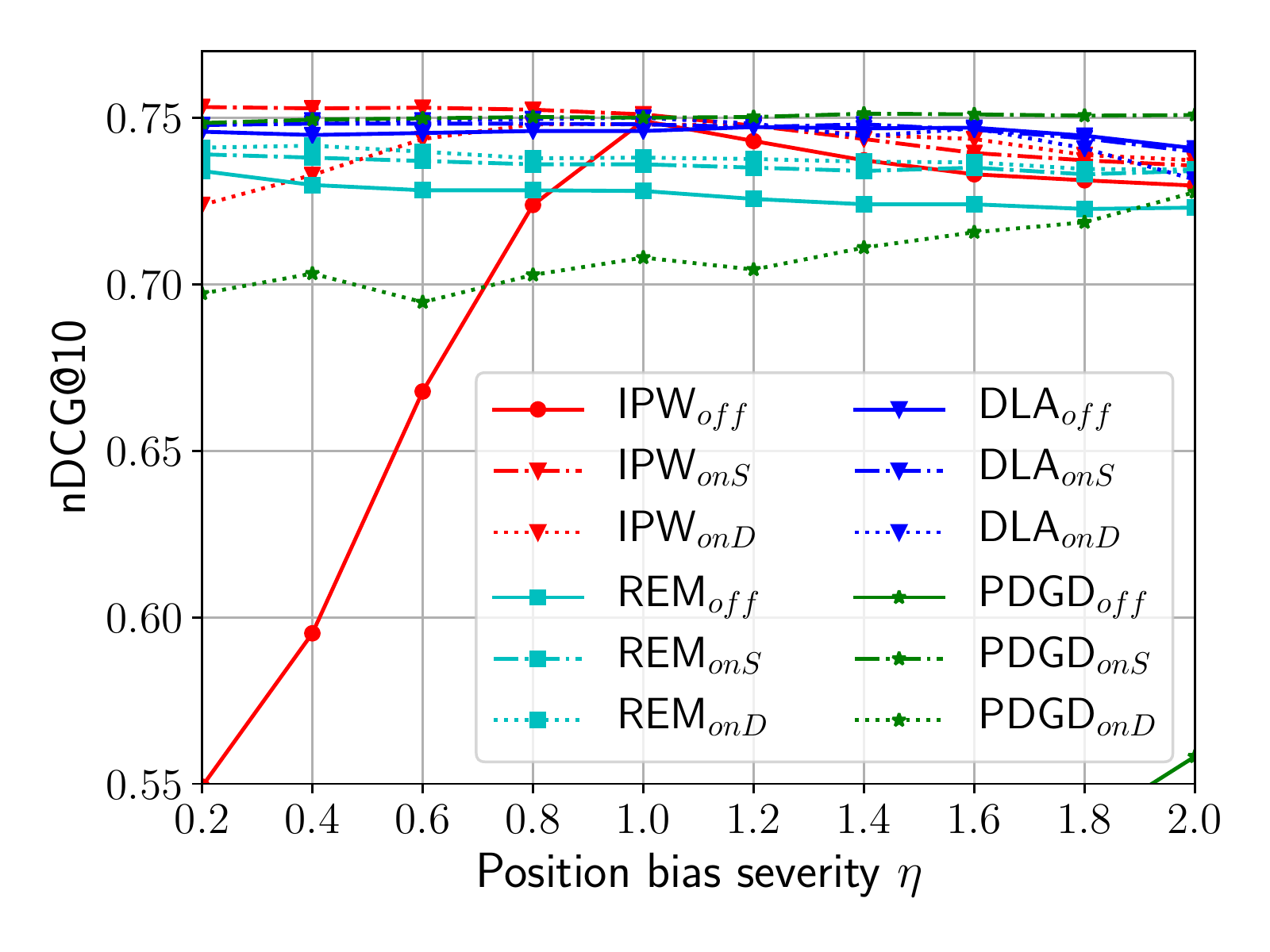}
		\caption{Ranking with linear regression}
		\label{fig:linear_eta}
	\end{subfigure}%
	\caption{Test performance with respect to bias severity $\eta$ on Yahoo! LETOR data.}
	\label{fig:eta}%
\end{figure}

To test \textbf{H3}, we repeated the same experiments on the Yahoo! dataset but changed the bias severity parameter $\eta$ to explore how the change of position bias would affect the performance of different unbiased learning-to-rank algorithms.
Figure~\ref{fig:eta} shows the nDCG@10 of different algorithms with $\eta$ from 0.2 (i.e., minor position bias) to 2.0 (i.e., severe position bias).
Because the inverse propensity weights of IPW is estimated with online result randomization experiments using click simulation with $\eta=1.0$, we observe that the IPW with offline learning (IPW$_{off}$) performed badly when $\eta$ is far from 1.0.
In contrast, the performance of other algorithms is relatively stable when $\eta$ changed from 0.2 to 2.0 because they either estimate examination propensity directly from the click data (e.g., DLA and REM) or manually exclude position bias from their training data by manipulating the distribution of the displayed ranked list (e.g., PDGD).
Particularly, the PDGD with stochastic online learning performed slightly better than other algorithms when the position bias is extremely low (i.e., $\eta=0.2$) or high (i.e., $\eta=2.0$).
This supports the hypothesis that, with proper learning paradigms, bandit learning algorithms are more robust to variable click bias (i.e., \textbf{H3}).


\subsection{Experiments with Real Data}\label{sec:real_exp}


In this section, we want to shed some lights on the actual performance of different unbiased learning-to-rank algorithms on real click data.
Due to the limit of our experiment resources, we are prohibited to do any types of online learning or online result randomization on real web search engines, so we focus on analyzing the performance of unbiased learning-to-rank algorithms in offline settings.
Specifically, we conduct offline experiments with the Tiangong dataset\footnote{\url{http://www.thuir.cn/data-tiangong-ultr/}}.
Tiangong dataset contains 3,449 queries with 3 million search sessions sampled from real search engine traffic as well as the top 10 documents and clicks.
Each query-document pair is represented with 33 standard ranking features extracted based on the term statistics, BM25, and language modeling scores on urls, titles, and document content.
A summary of the ranking features in Tiangong is shown in Table~\ref{tab:features}.
Also, Tiangong provides a separate test set with 100 queries and corresponding top 100 documents with 5-level relevance annotations for evaluation purpose.
To the best of our knowledge, this is the only public dataset that contains both user click data and human annotated relevance judgements.

\subsubsection{Model setup and evaluation}
Most settings of ranking models and loss functions are same to our experiments on synthetic data.
However, because the ranking features in Tiangong are highly limited, the performance of a ranking model could be severely affected by parameter initialization.
To guarantee the fairness of algorithm comparisons and the reproducibility of the experiments, we initialize all model parameters with a constant (i.e., 0.001).
Also, we reduced the hidden layer sizes of MLP to 64 and 32, and tuned the learning rate from 0.0001 to 0.005.
We used nDCG as our ranking metrics and reported the best test performance of each model after training.
Numbers are averaged from 10 repeated experiments to guarantee their credibility.
We ignore IPW and DBGD in this experiment because they require online result manipulations for estimating examination propensity or unbiased relevance signals.

\begin{table}[t]
	\caption{A summary of the ranking features in the Tiangong dataset.}
	\small
	\def\arraystretch{1.15}
	\begin{tabular}
		{| l | p{0.85\textwidth}|} \hline
		TF & The average term frequency of query terms in url, title, content and the whole document.   \\\hline
		IDF & The average inverse document frequency of query terms in url, title, content and the whole document.   \\\hline
		TF-IDF & The average value of $tf\cdot idf$ of query terms in url, title, content and the whole document.   \\\hline
		BM25 & The scores of BM25~\cite{robertson1994some} on url, title, content and the whole document.   \\\hline
		LMABS & The scores of Language Model (LM)~\cite{ponte1998language} with absolute discounting~\cite{zhai2017study} on url, title, content and the whole document.   \\\hline
		LMDIR & The scores of LM with Dirichlet smoothing~\cite{zhai2017study} on url, title, content and the whole document.   \\\hline
		LMJM & The scores of LM with Jelinek-Mercer~\cite{zhai2017study} on url, title, content and the whole document.   \\\hline
		Length & The length of url, title, content and the whole document.   \\\hline
		Slash & The number of slash in url.   \\\hline
	\end{tabular}\label{tab:features}
\end{table}

\subsubsection{Results}

Our experiment results on the Tiangong dataset are summarized in Table~\ref{tab:real}.
Here we report both the mean and the standard deviation of the 10 repeated runs for each algorithm.
As shown in the table, our experiment results on Tiangong are quite different from those on the synthetic data.
First, the performance of the naive algorithm (NA) that trains the ranking model with clicks directly is highly competitive.
As discussed previously, training ranking models with biased clicks are essentially optimizing the original rankings of documents. 
Because the original rankings of documents in Tiangong is created by the production system of the commercial Web search engine that have hundreds of features (much more than those released in the dataset), optimizing the original rankings would already produce good results.
Second, on Tiangong, REM performed the worst among all unbiased learning-to-rank algorithms.
Because the released features in Tiangong are simple text-matching features with limited expressive power, estimating relevance with pointwise loss functions using the 33 features on Tiangong is much more risky than using the 700 or 220 production features on the Yahoo! and Istella-S dataset.
Thus, it is reasonable to observe that REM, which uses the sigmoid pointwise loss, obtained worse performance than other models that use pairwise loss.

Similar to those observed on the synthetic data, DLA achieved the best performance among all unbiased learning-to-rank algorithms tested in our experiments in terms of NDCG.
Different from PDGD, DLA is theoretically guaranteed to find the unbiased ranking model in offline learning, and this advantage in theory has been reflected in the empirical experiments.
Besides, we observe that PairD also performed well on Tiangong and outperformed NA on several metrics.
Besides, we observe that PairD also performed well on Tiangong and outperformed NA on several metrics.
This indicates that PairD is capable of removing click bias in certain degree even though it's not theoretically principled.

\begin{table}[t]
	\caption{Comparison of unbiased learning-to-rank (ULTR) algorithms on Tiangong data. Numbers are shown with their standard deviation in 10 repeated runs (i.e., $\pm x$). 
	}
	\scalebox{0.83}{
	\begin{tabular}{  c || c | c | c | c | c | c | c | c   } \hline 
		Algorithms & nDCG@1 & ERR@1 & nDCG@3 & ERR@3 & nDCG@5 & ERR@5 & nDCG@10 & ERR@10 \\ \hline \hline
		REM & 0.444$_{\pm .011}$ & 0.416$_{\pm .0100}$ & 0.437$_{\pm .011}$ & 0.558$_{\pm .0063}$ & 0.439$_{\pm .010}$ & 0.587$_{\pm .005}$ & 0.452$_{\pm .004}$ & 0.599$_{\pm .005}$ \\\hline
		DLA & 0.453$_{\pm .005}$ & 0.425$_{\pm .0046}$ & 0.460$_{\pm .002}$ & 0.562$_{\pm .002}$ & \textbf{0.458}$_{\pm .002}$ & 0.589$_{\pm .003}$ & \textbf{0.471}$_{\pm .001}$ & 0.604$_{\pm .002}$ \\\hline
		PairD & 0.445$_{\pm .010}$ & 0.418$_{\pm .0095}$ & 0.448$_{\pm .009}$ & \textbf{0.565}$_{\pm .010}$ & 0.453$_{\pm .005}$ & \textbf{0.591}$_{\pm .009}$ & 0.465$_{\pm .002}$ & \textbf{0.605}$_{\pm .008}$ \\\hline \hline
		PDGD & 0.433$_{\pm .008}$ & 0.402$_{\pm .0080}$ & 0.430$_{\pm .004}$ & 0.553$_{\pm .0042}$ & 0.443$_{\pm .003}$ & 0.584$_{\pm .005}$ & 0.464$_{\pm .001}$ & 0.596$_{\pm .004}$ \\\hline \hline
		NA & \textbf{0.455}$_{\pm .006}$ & \textbf{0.427}$_{\pm .0055}$ & \textbf{0.461}$_{\pm .004}$ & 0.563$_{\pm .003}$ & 0.455$_{\pm .002}$ & 0.589$_{\pm .003}$ & 0.467$_{\pm .001}$ & 0.603$_{\pm .003}$ \\\hline

	\end{tabular}
	}
	\label{tab:real}
\end{table}





\section{Conclusion}\label{sec:conclusion}

In this paper, we discuss the differences and connections between unbiased learning to rank algorithms proposed in offline settings and online settings.
We show that the existing studies on the counterfactual learning algorithms and the bandit learning algorithms are essentially solving the same problem from two theoretical perspectives.
We formally evaluate eight state-of-the-art unbiased learning to rank algorithms and show how different offline and online learning paradigms would affect the theoretical foundations and empirical effectiveness of each algorithm.



As demonstrated in Section~\ref{sec:theory} and \ref{sec:exp}, the unbiasness of counterfactual learning algorithms are invariant to the distribution of the displayed results while the unbiasness of bandit learning algorithms are more robust to the variance of click bias.
Whether these properties benefit or hurt their applications in practice varies from cases to cases.
For example, when user satisfaction is not sensitive to the quality of result ranking and we have full control over the result pages, bandit learning algorithms may produce more robust ranking models than counterfactual learning through real-time interactions with end users.
However, in search engines where no single model can fully control the final displayed ranked lists (e.g., the result pages of most commercial search engines and recommendation systems are the combination of ads and organic results produced by multiple models), counterfactual learning is preferable to bandit learning because the effectiveness of the later is extremely sensitive to the final distribution of the displayed result lists.





\section{Acknowledgments}
This work was supported by the School of Computing, University of Utah. Any opinions, findings and conclusions or recommendations expressed in this material are those of the authors and do not necessarily reflect those of the sponsor.

\bibliographystyle{ACM-Reference-Format-Journals}
\bibliography{sigproc} 

\newpage

\end{document}